\documentclass[twocolumn,prd,superscriptaddress,preprintnumbers,amsmath,amssymb,nofootinbib]{revtex4-1}

\usepackage{amsmath}
\usepackage{amssymb}
\usepackage{slashed}
\usepackage{graphicx}
\usepackage{graphics}
\usepackage{epsfig}
\usepackage{color}
\usepackage{hyperref}
\newcommand{\Eqref}[1]{Eq.~\eqref{#1}}
\newcommand{\bare}{\bar{e}}

\newcommand{\Nf}{N_{\text{f}}}
\newcommand{\pat}{\partial_t}

\begin{document}
\title{Light fermions in quantum gravity}

\author{Astrid Eichhorn and Holger Gies}
\affiliation{\mbox{\it Theoretisch-Physikalisches Institut, Friedrich-Schiller-Universit{\"a}t Jena,}
\mbox{\it Max-Wien-Platz 1, D-07743 Jena, Germany}
\mbox{\it E-mail: {astrid.eichhorn@uni-jena.de, holger.gies@uni-jena.de}}
}
\begin{abstract} 
  We study the impact of quantum gravity, formulated as a quantum field theory
  of the metric, on chiral symmetry in a fermionic matter sector. We
  specifically address the question as to whether metric fluctuations can
  induce chiral symmetry breaking and bound state formation. Our results based
  on the functional Renormalization Group indicate that chiral symmetry is
  left intact even at strong gravitational coupling. In particular, we find
  that asymptotically safe quantum gravity where the gravitational couplings
  approach a non-Gau\ss ian fixed point generically admits universes with
  light fermions. Our results thus further support quantum gravity theories
  built on fluctuations of the metric field such as the asymptotic-safety
  scenario.  A study of chiral symmetry breaking through gravitational quantum
  effects may serve as a significant benchmark test also for other quantum
  gravity scenarios, since a completely broken chiral symmetry at the Planck
  scale would not be in accordance with the observation of light fermions in
  our universe.  We demonstrate that this elementary observation already
  imposes constraints on a generic UV completion
  of gravity.
\end{abstract}

\maketitle

\section{Introduction}

Any phenomenologically relevant theory of quantum gravity has to satisfy a
number of physical requirements. In addition to internal or mathematical
consistency, observations demand that such a theory has to provide for the
existence of a semi-classical limit. Another phenomenological requirement is
the existence of light matter (compared with the Planck scale). 
In many studies of quantum gravity, matter is either ignored, or treated
rather as a kinematic degree of freedom. For instance in string
compactifications, information about the existence of light matter is drawn
from anomaly-cancelation arguments.

As soon as matter and interactions are taken into account dynamically, the
existence of light matter is far from being self-evident: From a bottom-up
viewpoint, the regime of quantum gravity is naturally defined as the domain
where gravity fluctuations become as relevant as matter fluctuations. If
gravity becomes even strongly interacting, its dynamical influence on the
matter sector may even be similar to strong matter correlations as
induced by the other forces of particle physics. 

This question becomes particularly paradigmatic in the case of fermions. In
standard particle physics scenarios, fermions are light because their mass is
protected by chiral symmetry. So far, the mass of all fermionic matter content
in the universe is associated with the phenomenon of chiral symmetry
breaking. This mass generation is a consequence of strong correlations among
fermions. In the electroweak sector, these strong correlations are provided by
the Higgs sector, whereas gluon-induced interactions are responsible for mass
generation in the strong interactions. 

Gravity at first sight seems to have some similarity to these particle physics
examples, as it is similar to Yang-Mills theories in many respects. Therefore,
the present work is devoted to a first investigation whether gravitational
fluctuations can induce chiral symmetry breaking in a chiral fermion
sector. If such a mechanism exists, any theory of quantum gravity would have
to control it or evade it in a natural way. Since such a mechanism would be
active at or above the Planck scale, it would naturally force fermions to have 
masses on the order of the Planck scale in contrast to observation. 

In our work, gravitational fluctuations are parameterized as fluctuations of
the metric field. This is at least an effective description below or even near
the Planck scale. In the context of Weinberg's asymptotic safety scenario for
quantum gravity \cite{Weinberg:1980gg}, this can even be a consistent
description up to arbitrarily short distances.
As asymptotic safety is built on the existence of  a non-Gau\ss ian fixed
point in the gravitational couplings, i.e., an interacting UV limit, the
interplay between the gravitational sector and the chiral fermion sector is of
particular interest. 

Evidence for the existence of such a fixed point has been provided in
different approaches \cite{Gastmans:1977ad,Christensen:1978sc,Smolin:1981rm,
  Niedermaier:2010zz, Hamber:2009zz}. Also causal dynamical triangulations
(see e.g.  \cite{Ambjorn:2010rx} for a review) may be interpreted as providing
evidence for this scenario.  In particular the functional
Renormalization
Group (RG), following the pioneering work of Reuter \cite{Reuter:1996cp} has
facilitated numerous studies supporting asymptotic safety in gravity
\cite{Dou:1997fg,Souma:1999at,Lauscher:2001cq,Lauscher:2001ya,Lauscher:2001rz,Reuter:2001ag,Lauscher:2002sq,Reuter:2002kd,Litim:2003vp,Bonanno:2004sy,Lauscher:2005xz,Reuter:2005bb,Fischer:2006fz,Codello:2006in,Codello:2007bd,Machado:2007ea,Codello:2008vh,Reuter:2008wj,Benedetti:2009rx,Eichhorn:2009ah,Eichhorn:2010tb,Groh:2010ta,
  Manrique:2010am,Percacci:2010yk,Daum:2010qt,Manrique:2011jc}, for reviews see
\cite{Niedermaier:2006wt,Niedermaier:2006ns,Percacci:2007sz,reviews_AS}. First
steps in the investigation of the
compatibility of asymptotically safe quantum gravity with
quantized matter have been performed in
\cite{Percacci:2002ie, Percacci:2003jz}. Here the
backreaction of fermionic and bosonic matter onto the
gravitational fixed-point properties were investigated. 
The requirement of the existence of a physically admissible
fixed point  with a positive value for the Newton coupling then
imposes constraints on the matter content of the universe.
Most importantly, the matter content of the standard model
of particle physics is compatible with asymptotically safe
quantum gravity within the investigated truncation
\cite{Percacci:2002ie}. (For work on the effect of the
gravitational fixed point on gauge theories see
\cite{Daum:2009dn,Daum:2010bc,
Harst:2011zx,Folkerts:2011jz}.
Studies dealing with a possible solution of the
triviality problem in the Higgs sector through the coupling
to gravity have been performed in
\cite{Zanusso:2009bs, Shaposhnikov:2009pv, Vacca:2010mj}. 
)

In this work, we explore the gravity-induced interactions and potentially
strong correlations in a chiral fermion sector for the first time using the
functional RG.  As gravity shares some features with Yang-Mills theories, we
are motivated by recent advances in QCD, where chiral symmetry breaking can be
understood as the consequence of a critical dynamics among the chiral fermions
which is triggered by gluon-induced strong correlations
\cite{Gies:2002hq,Gies:2005as}; for a successful determination of the critical
temperature for chiral symmetry breaking in QCD using the functional RG, see
\cite{Braun:2005uj,Braun:2006jd,Braun:2009gm,Kondo:2010ts}. It is tempting to
speculate that gravity might facilitate a similar mechanism in a
strong-coupling regime. If so, such a mechanism might exhibit a dependence on
control parameters of the theory such as the fermion flavor number $\Nf$. In
fact, chiral quantum phase transitions as a function of $\Nf$ have been
observed in many systems such as many flavor QCD, QED$_3$ or the 3-dimensional
Thirring model
\cite{Banks:1981nn,Miransky:1996pd,Appelquist:1996dq,Gies:2005as,Braun:2010qs,%
  Pisarski:1984dj,Fischer:2004nq,Christofi:2007ye,Gies:2010st}.

Our study of the gravitationally-stimulated chiral dynamics is based on a
truncation of the full quantum effective action that concentrates on a
Fierz-complete basis of chiral fermionic four-point functions in the
point-like limit. Again, this is motivated by analogous studies in other
theories, where such an ansatz provides both for an intuitive as well as
quantitatively meaningful approach to chiral symmetry breaking. 

As our main result, we do not find any indications for
gravitationally-stimulated chiral symmetry breaking within this
ansatz. Whereas the Gau\ss ian fermion matter fixed point turns into an
interacting non-Gau\ss ian one, the universality properties of this fixed
point receive rather small modifications in the asymptotic-safety scenario
if fermionic degrees of freedom dominate the matter sector. As a
general pattern, gravitational binding which would favor chiral
  symmetry breaking is compensated by gravitational contributions to anomalous
  scaling of the fermion interactions. Within this minimal truncation, we can 
therefore conclude that asymptotic safety is well compatible with the
existence (and observation) of light fermions despite an interacting UV sector
which stimulates fermion self-interactions.

This paper is structured as follows: We will introduce the
functional RG as a tool for our investigation in
Sect.~\ref{Wetteq}, and introduce the system that
we study in Sect.~\ref{truncation}. Results concerning the asymptotic-safety
scenario as well as for general classes of effective theories of quantum
gravity are presented in Sect.~\ref{results}. Finally, we conclude in
Sect.~\ref{conclusions}. Technical details can be found in
appendix \ref{appendix}.

\section{Functional RG}\label{Wetteq}
The functional RG facilitates the
 non-perturbative evaluation of full correlation functions. Within
the formulation used here, we focus on the scale-dependent
effective average action which is the generating functional of 1PI correlators
that include all fluctuations from the ultraviolet (UV) down to the infrared
(IR) scale $k$. At $k=0$, $\Gamma_k$ coincides with the standard effective
action $\Gamma=\Gamma_{k=0}$. The scale dependence of the effective
average action is governed by the Wetterich equation
\cite{Wetterich:1993yh}
\begin{equation}
 \partial_t \Gamma_k = \frac{1}{2}{\rm STr} \{
 [\Gamma_k^{(2)}+R_k]^{-1}(\partial_t R_k)\}.
\end{equation}
Here, $\partial_t = k \,\partial_k$, $\Gamma_k^{(2)}$ is the second functional
derivative of $\Gamma_k$ with respect to the fields, and $R_k$ is an IR
regulator function. Accordingly the right-hand side of the Wetterich equation
depends on the full (field-dependent) regularized propagator
$\left(\Gamma_k^{(2)}+R_k \right)^{-1}$, which is matrix-valued in field
space. The supertrace $\rm STr$ contains a trace over the spectrum of the full
propagator in all appropriate indices (i.e. on a flat background in the
absence of classical background fields it translates into a momentum integral
and a trace over Lorentz and internal indices). For Grassmann-valued fields
the supertrace involves an additional negative sign.  For reviews on the
functional RG and the Wetterich equation see
\cite{Berges:2000ew,Polonyi:2001se,
  Pawlowski:2005xe,Gies:2006wv,Delamotte:2007pf,Rosten:2010vm}.

Since quantum fluctuations generate all possible operators compatible with the
symmetries and the field content of the microscopic action, the effective
(average) action lives in theory space which is spanned by all these
operators. Expanding $\Gamma_k= \sum_n g_n(k) \mathcal{O}_n$ into the infinite
sum of all operators $\mathcal{O}$ with running couplings $g_n(k)$ allows to
rewrite the Wetterich equation as an infinite tower of coupled differential
equations. In practice, a truncation of theory space to a smaller subspace is
necessary.  In the case of asymptotically safe quantum gravity, numerous
studies have shown a high degree of stability of truncations of the
Einstein-Hilbert type and beyond. The results appear to converge under generalizations of the
truncation in various ways, as well as under a change of the regularization
scheme, 
thus providing strong support for the existence of
the NGFP in full theory space \cite{Reuter:1996cp, Dou:1997fg,Souma:1999at,Lauscher:2001cq,Lauscher:2001ya,Lauscher:2001rz,Reuter:2001ag,Lauscher:2002sq,Reuter:2002kd,Litim:2003vp,Bonanno:2004sy,Lauscher:2005xz,Fischer:2006fz,Codello:2006in,Codello:2007bd,Machado:2007ea,Codello:2008vh,Reuter:2008wj,Benedetti:2009rx,Eichhorn:2009ah,Eichhorn:2010tb,Groh:2010ta,
  Manrique:2010am,Manrique:2011jc}.

The perturbative non-unitarity of theories involving
higher derivative operators does not directly apply here:
Indeed unitarity has to be reinvestigated within the
non-perturbative setting and can remain intact within the
asymptotic-safety scenario (see e.g. corresponding discussions in
\cite{Lauscher:2002sq,Niedermaier:2006wt, Percacci:2007sz,Benedetti:2009rx}).

In the following, we investigate the compatibility of the asymptotic-safety
scenario with the existence of light fermionic matter. More generally, our
results can be applied to generic effective theories of quantum gravity
formulated in terms of fluctuations of the metric field and their interplay
with chiral fermions. 

\section{Chiral fermions in quantum Einstein gravity}\label{truncation}

In the case of QCD-like theories, many studies based on functional methods
suggest that chiral symmetry is broken for gauge couplings larger than a
critical value
\cite{Miransky:1988gk,Aoki:1999dv,Aoki:1999dw,Gies:2002hq,Gies:2005as}. In
direct analogy it is tempting to expect that there exists a critical value for
the Newton coupling at which metric fluctuations break chiral symmetry.  This
would agree with the picture that gravity is always attractive and thus should
support fermionic binding phenomena.

We investigate this scenario in a
specific fermionic system with a chiral SU$({\rm N_f})_{\mathrm{L}}$ $\times$
SU$({\rm N_f})_{\mathrm{R}}$ symmetry. We parameterize this system by an
action of the form
\begin{eqnarray}
 \Gamma_{k\, \rm F}&=& \int d^4x \sqrt{g}\, i Z_{\psi}
\bar{\psi}^i
\gamma^{\mu}\nabla_{\mu}\psi \label{eq:GammaF}\\
&& + \frac{1}{2} \int d^4x \, \sqrt{g}\,
 \bigl[\bar\lambda_-(k) (V-A) + \bar\lambda_+(k)
(V+A ) \bigr],\nonumber
\end{eqnarray}
where 
\begin{eqnarray}
 V&=& \left( \bar{\psi}^i \gamma_{\mu}\psi^i \right)\left( \bar{\psi}^j
   \gamma^{\mu}\psi^j \right), \\
A&=&- \left( \bar{\psi}^i \gamma_{\mu}\gamma^5\psi^i \right)\left(
  \bar{\psi}^j \gamma^{\mu}\gamma^5\psi^j \right). 
\end{eqnarray}
These fermionic self-interactions form a complete basis of four-fermion
operators in the pointlike limit which are invariant under the chiral
symmetry. The parentheses indicate expressions with fully
 contracted Dirac indices. The $\gamma$ matrices are
understood to live in curved spacetime, being related to their
flat-space cousins $\gamma^a$ by the vielbein:
$\gamma^{\mu}= e^{\mu}_a \gamma^a$. Flavor indices are
denoted by Latin letters $i,j,...$ and run from 1 to $\rm
N_{f}$.
The covariant derivative $\nabla_{\mu}$ is given 
by $\nabla_{\mu}\psi= \partial_{\mu}\psi +
\frac{1}{8}[\gamma^a,\gamma^b]\omega_{\mu\, ab}\psi$, where
$\omega_{\mu}^{ab}$ denotes the spin connection. The latter
can be determined in terms of the Christoffel
connection by demanding that $\nabla_{\mu}e^{\mu}_a =0$.

All other non-derivative SU$({\rm N_f})_{\mathrm{L}}$ $\times$ SU$({\rm
  N_f})_{\mathrm{R}}$ symmetric four-fermion operators, e.g. a flavor
non-singlet scalar-pseudo-scalar interaction, can be
transformed into some
combination of the above ones by a Fierz transformation. Including all of the
basis operators implies that we cover all possible channels for chiral
symmetry breaking in the point-like, i.e. momentum-independent, limit. This is
important, as gravity might pick one specific channel to induce the breaking
of chiral symmetry.
This ansatz of operators in the chiral sector is strongly motivated by similar
lines of reasoning in QCD-like theories or other strongly-correlated fermionic
systems. There, the ansatz is capable of describing the
approach to chiral
symmetry breaking qualitatively as well as quantitatively. Of course, 
gravity may choose to break chiral symmetry in a fashion differing from
Yang-Mills theory; potential further mechanisms will be briefly outlined
below. 

In $d>2$ dimensional spacetime, four-fermion interactions are perturbatively
non-renormalizable. In an RG language this translates into the fact that these
couplings are irrelevant at the Gau\ss{}ian fixed point. Accordingly they have
to be set to zero at the UV scale in a fundamental theory in a
  perturbative setting. Even if being zero
initially, such couplings are generated by interactions in the flow towards
the IR, for instance, in the context of QCD or also when coupled to
gravity. The flow of such fermionic self-interactions can then provide
indications for chiral symmetry breaking. Beyond perturbation theory,
non-Gau\ss{}ian fixed points may exist which could allow to construct a
non-perturbatively renormalizable (asymptotically safe) theory with
non-vanishing four-fermion interactions. In $2<d<4$
dimensions, for instance, the Gross-Neveu model provides for a simple and
well-understood example of asymptotic safety \cite{Braun:2010tt}. 

At this point, we can already discuss the relation between chiral symmetry
breaking and the fixed-point structure of the four-fermion couplings
$\bar\lambda_{\pm}$ \cite{Gies:2005as}. For this, we introduce the
dimensionless renormalized couplings $\lambda_\pm$ and the fermionic anomalous
dimension $\eta_\psi$:
\begin{equation}
\lambda_\pm = \frac{k^2 \bar\lambda_\pm}{Z_\psi}, \quad 
\eta_\psi = -\pat \ln Z_\psi. 
\label{eq:dimrenlambda}
\end{equation}
Due to the one-loop form of the Wetterich equation, the $\beta$ functions for
$\lambda_\pm$ have the generic form
\begin{equation}
 \beta_{\lambda_{\pm}}= (2+\eta_{\psi})\lambda_{\pm}+a \,\lambda_{\pm}^2 +b\, 
\lambda_{\pm}\lambda_{\mp} + c \,\lambda_{\mp}^2+d \lambda_{\pm}+e.
\label{eq:betalambda}
\end{equation}
Herein the first term arises from dimensional (and anomalous) scaling. The
quadratic contributions follow from a purely fermionic two-vertex diagram
(cf. diagram (2c) in Fig.~\ref{diagrams} below). A tadpole contribution $\sim
d\lambda_{\pm}$ may also exist, as well as a $\lambda_{\pm}$-independent part
$\sim e$ which results from the coupling to other fields, for instance,
arising from the covariant derivative in the kinetic term in
\Eqref{eq:GammaF}. The numerical values for $a,b$ and $c$ depend on the
regulator, the contributions $d$ and $e$ will also depend on further
couplings. Specific representations will be given below.

Fixing all other couplings, the $\beta$ function of a given fermionic coupling
$\beta_{\lambda_\pm}= \pat \lambda_\pm$ as a function of $\lambda_\pm$ is a
parabola with two fixed points $\lambda_\pm^{\ast}$ where
$\beta_{\lambda_\pm}(\lambda_\pm^\ast) =0$. The coupled system of two fermionic
$\beta$ functions then admits $2^2$ fixed points
which need not necessarily all be real.

\begin{figure}
\includegraphics[width=0.5\textwidth,height=0.3\textwidth]{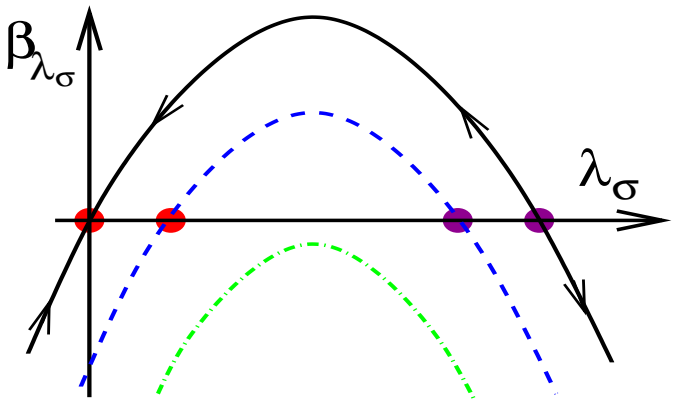}
\caption{Sketch of the $\beta$ function for the chiral channel
  $\lambda_\sigma$. In the absence of further interactions, the parabola-type
  $\beta$ function exhibits two fixed points, the Gau\ss ian fixed point at
  $\lambda_\sigma^\ast=0$ and a non-Gau\ss ian fixed point at
  $\lambda_\sigma^\ast=\lambda_{\sigma, \text{cr}}>0$. Arrows indicate the RG
  flow towards the IR. Further interactions can shift the parabola (dashed
  line) and lead to an annihilation of the two fixed points (dotdashed
  line). This annihilation is an indicator for an approach to chiral
  criticality. Such a scenario typically occurs in QCD-like theories
  \cite{Kondo:1991yk,Gies:2003dp,Gies:2005as,Braun:2009ns}.} 
\label{fig:parabolasketch}
\end{figure}

In order to illustrate the relevance of this fermionic fixed-point structure,
let us concentrate on the $\lambda_+$ channel and perform a Fierz
transformation to the standard scalar-pseudo-scalar channel,
\begin{equation}
\lambda_+ \big[(\bar\psi^i\gamma_\mu \psi^i)^2-(\bar\psi^i\gamma_\mu \gamma_5
  \psi^i)^2\big] 
= \lambda_\sigma\big[(\bar\psi^i \psi^j)^2-(\bar\psi^i\gamma_5
  \psi^j)^2\big],
\label{eq:Fierz}
\end{equation}
where $(\bar\psi^i \psi^j)^2 \equiv \bar\psi^i \psi^j \bar\psi^j \psi^i$, and
similarly for the pseudo-scalar channel. Equation \eqref{eq:Fierz} is an exact
Fierz identity if the couplings satisfy
\begin{equation}
\lambda_\sigma = -\frac{1}{2} \lambda_+.
\end{equation}
The fixed-point structure in the $(V+A)$ channel hence implies a corresponding
fixed-point structure in the standard scalar-pseudo-scalar channel, where
chiral symmetry breaking is expected to be visible. In
Fig.~\ref{fig:parabolasketch}, the $\beta$ function $\beta_{\lambda_{\sigma}}=\pat\lambda_\sigma$ for this chiral channel
 is sketched for vanishing gravitational coupling. The two
crossings of the parabola with the $\lambda_\sigma$ axis indicate the two
fixed points. The Gau\ss ian fixed point at $\lambda_\sigma^\ast=0$ is IR
attractive whereas the non-Gau\ss ian fixed point
$\lambda_\sigma^\ast=\lambda_{\sigma, \text{cr}}>0$ is UV
attractive (arrows indicate the flow towards the IR). If the system is in the
attractive domain of the Gau\ss ian fixed point, fermionic correlations die
out quickly during the RG flow and the system remains in the chirally
symmetric phase. 

If the system starts to the right of the non-Gau\ss ian fixed point
$\lambda_\sigma > \lambda_{\sigma, \text{cr}}$, the coupling runs towards very
large values and actually diverges at a finite RG scale. Whereas the
point-like fermionic truncation breaks down here, a transition to a bosonized
description 
with a boson being related to a fermion bilinear
identifies the chiral coupling with the mass term of a chiral
bosonic field $\lambda_\sigma \sim 1/m_\phi^2$. The divergence of the
fermionic coupling therefore indicates that the bosonic mass term drops below
zero which is a signature for symmetry breaking. It characterizes the onset of
the effective chiral potential to acquire a Mexican-hat form. In NJL-type
models, the non-Gau\ss ian fixed point is identical to the critical value of
the coupling $\lambda_{\sigma, \text{cr}}$ required to put the system into the
chiral symmetry broken phase. 

A new feature comes about by coupling the fermionic system to another
interaction, e.g. a gauge theory. Then, an originally weakly coupled fermion
sector near the Gau\ss ian fixed point can dynamically be driven to chiral
criticality. For this, the Gau\ss ian fixed point needs to annihilate with the
non-Gau\ss ian fixed point, such that the corresponding $\beta$ function for
$\lambda_\sigma$ becomes completely negative, as indicated by the dashed and
dotdashed line in Fig.~\ref{fig:parabolasketch}. Fixed-point annihilation thus
is an indicator for an approach to chiral criticality. This scenario is indeed
realized, e.g. in QCD-like and other theories
\cite{Kondo:1991yk,Gies:2003dp,Gies:2005as,Braun:2009ns}. With
regard to the structure of \Eqref{eq:betalambda}, the last two terms together
with the gravitational contribution to $\eta_\psi$, in principle, have the
potential to destabilize the fermionic fixed points and drive the fermionic
system to criticality. Whether or not this happens is a quantitative question
that has to be addressed by integrating out gravitational fluctuations in a
given quantum gravity theory.

For the gravitational part we work in the background field
 formalism \cite{Abbott:1980hw}, where the full metric is
split according to
\begin{equation}
 g_{\mu \nu}= \bar{g}_{\mu \nu}+ h_{\mu \nu},
\end{equation}
where this split does not imply that we consider only small fluctuations
around e.g. a flat background. Within the FRG approach we have access to
physics also in the fully non-perturbative regime.  This formalism, being
highly useful in non-abelian gauge theories (see, e.g. 
\cite{Pawlowski:2005xe,Gies:2006wv}), is mandatory in gravity, since the
background metric allows for a meaningful notion of "high-momentum" and
"low-momentum" modes as implied by the spectrum of the background covariant
Laplacian.

The desired feature of background independence naively seems spoiled in this
way; however, the background formalism turns out to be merely a technical
tool, leaving physical results unaffected and thus independent of the
background \cite{Niedermaier:2006ns}.

Following standard approximations we work within a
single-metric truncation, i.e. we set $g_{\mu \nu}=
\bar{g}_{\mu \nu}$ after evaluating $\Gamma_k^{(2)}$. As
suggested in recent studies \cite{Manrique:2010am,
Manrique:2010mq} qualitative results in the Einstein-Hilbert
sector are not affected by this approximation.

On a general (curved) spacetime our truncation then reads:
\begin{eqnarray}
 \Gamma_k = \Gamma_{k\, \rm EH}+ \Gamma_{k\,\rm gf}
+ \Gamma_{k\, \rm F},
\end{eqnarray}
where the Einstein-Hilbert term and the gauge-fixing term are given by:
\begin{eqnarray}
\Gamma_{k\,\mathrm{EH}}&=& 2 \bar{\kappa}^2 Z_{\text{N}} (k)\int 
d^4 x \sqrt{g}(-R+ 2 \bar{\lambda}(k))\label{eq:GEH},\\
\Gamma_{k\,\mathrm{gf}}&=& \frac{Z_{\text{N}}(k)}{2\alpha}\int d^4 x
\sqrt{\bar g}\, \bar{g}^{\mu \nu}F_{\mu}[\bar{g}, h]F_{\nu}[\bar{g},h]\label{eq:Ggf},
\end{eqnarray}
with
\begin{equation}
 F_{\mu}[\bar{g}, h]= \sqrt{2} \bar{\kappa} \left(\bar{D}^{\nu}h_{\mu
   \nu}-\frac{1+\rho}{4}\bar{D}_{\mu}h^{\nu}{}_{\nu} \right). 
\end{equation}
Herein, $\bar{\kappa}= (32 \pi G_{\text{N}})^{-\frac{1}{2}}$ is related to the
bare Newton constant $G_{\text{N}}$. We denote
the cosmological constant by $\bar\lambda$ without any
subscript. It should not be confused with the four-fermion couplings
$\bar\lambda_{\pm}$.

A ghost sector corresponding to Faddeev-Popov gauge fixing is implicitely
understood here. 

For the vielbein we work in the symmetric vielbein 
gauge \cite{Woodard:1984sj, vanNieuwenhuizen:1981uf} such
that $O(4)$ ghosts do not occur. This gauge also allows to
reexpress vielbein fluctuations purely in terms of metric
fluctuations.

Details on the second functional derivative of the effective
action can be found in appendix \ref{appendix}.

The fermionic self-interactions $\sim \bar\lambda_\pm$ do not directly
contribute to the pure gravity flow. Technically this is, because no one-loop
diagram containing a fermionic four-point vertex can be formed that has only
gravitons on external legs. Hence, the Einstein-Hilbert sector receives
contributions only from the minimally coupled kinetic fermion term as
determined in \cite{Percacci:2002ie}, where the approximation $Z_{\psi}=1$ has
been used.

Our new task here is to compute the gravitationally
stimulated flow of the
fermion interactions $\bar\lambda_\pm$. For this, a 
flat-background calculation, setting 
$\bar{g}_{\mu \nu}= \delta_{\mu \nu}$, is fully sufficient,
and technically favorable.

In the following, we ignore a non-trivial running of the fermion kinetic term
by setting $Z_\psi=1$. In Yang-Mills theory, this is justified in the Landau
gauge $\alpha\to 0$, where this flow of $Z_\psi$ vanishes in a similar
truncation \cite{Gies:2003dp}. In gravity, however, the flow of $Z_\psi$ does
receive non-trivial contributions even in the Landau-deWitt gauge
$\rho\to\alpha\to0$. This marks a first difference between gravity and
Yang-Mills theory in this context. We keep track of this difference by
maintaining the dependence of the flow on the fermion anomalous dimension
$\eta_\psi$.  Whereas $\eta_\psi=0$ in the present approximation, we will
later treat $\eta_\psi$ as a free parameter to explore possible consequences
of this difference between Yang-Mills theory and gravity.

We decompose the metric fluctuations $h_{\mu \nu}$ into a
transverse traceless tensor, a transverse vector, a scalar,
and the trace part. We then specialize to Landau deWitt
gauge, $\rho \rightarrow \alpha \rightarrow 0$, which
implies that only the transverse traceless tensor $h_{\mu
\nu}^{\mathrm{TT}}$ and the trace mode $h= \bar{g}^{\mu
\nu}h_{\mu \nu}$ can contribute to the flow of the fermionic
couplings (see also \cite{Eichhorn:2010tb}).

Splitting $\Gamma_k^{(2)}+R_k =
\mathcal{P}_k+\mathcal{F}_k$, where all field-dependent
terms enter the fluctuation matrix $\mathcal{F}_k$, we may
now expand the right-hand side of the flow equation as
follows:
\begin{eqnarray}
 \partial_t \Gamma_k&=& \frac{1}{2}{\rm STr} \{
 [\Gamma_k^{(2)}+R_k]^{-1}(\partial_t R_k)\}\label{eq:flowexp}\\
&=& \frac{1}{2} {\rm STr}\, \tilde{\partial}_t\ln
\mathcal{P}_k
+\frac{1}{2}\sum_{n=1}^{\infty}\frac{(-1)^{n-1}}{n} {\rm
  STr}\,
\tilde{\partial}_t(\mathcal{P}_k^{-1}\mathcal{F}_k)^n,
\nonumber
\end{eqnarray}
where the derivative $\tilde{\partial}_t$ in the second line by definition
acts only on the $k$ dependence of the regulator, $\tilde{\partial}_t=
\int \partial_t R_k\frac{\delta}{\delta R_k}$. Since each
factor of $\mathcal {F}_k$ contains a
coupling to external fields, this expansion simply corresponds to an expansion
in the number of vertices. As we are interested in the flow
of the four-fermion coupling, we can neglect terms with
more than four vertices. The contributing
terms are then given by the diagrams in Fig.~\ref{diagrams}.

Diagrams (2c), (3a), (4a), and (4b) occur when Yang-Mills theory is coupled to
fermions minimally.  The additional diagrams can be traced back to the fact
that the volume element containing $\sqrt{g}$ generates additional
graviton-fermion-couplings. Also the covariant derivative in the kinetic term
generates not only one- but also two-graviton fermion couplings. An additional
triangular diagram, built from a two-fermion-two-graviton vertex and two
vertices coupling the graviton to one external and one internal (anti)fermion,
vanishes in the Landau-deWitt gauge. The former vertex exists only for the
transverse traceless, whereas the latter couples only to the trace mode.
Since the metric propagator is diagonal in these modes for all choices of gauge parameters $\alpha$ and $\rho$, a non-vanishing
diagram of this type cannot be constructed.
\begin{figure}
\setlength{\unitlength}{1cm}
\begin{picture}(8,12)
\put(0.5,8){\includegraphics[scale=0.5]{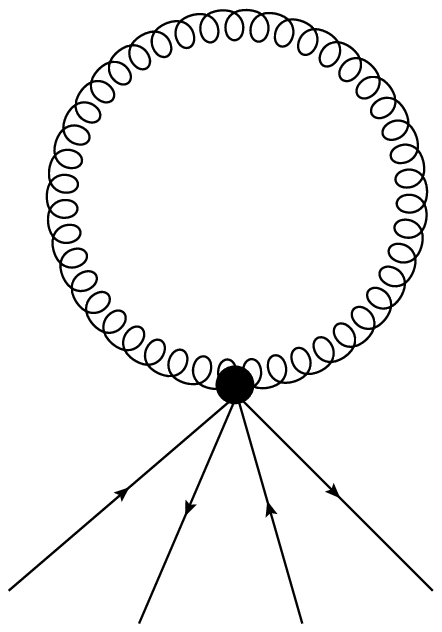}}
\put(2.67,9.1){\circle{0.5}}
\put(2.5,9){1a}
\put(4.7,8){\includegraphics[scale=0.5]{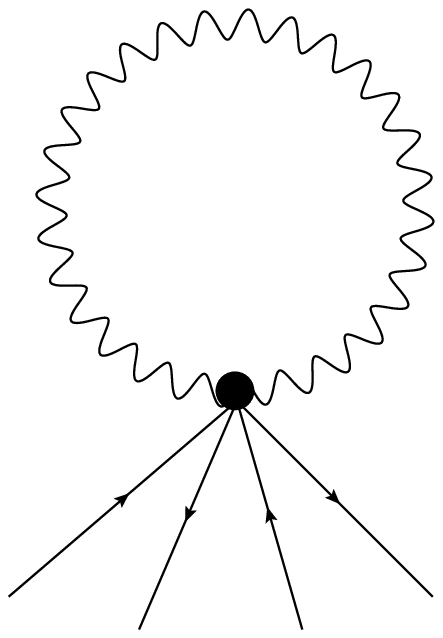}}
\put(6.87,9.1){\circle{0.5}}
\put(6.7,9){1b}
\put(0,6){ \includegraphics[scale=0.5]{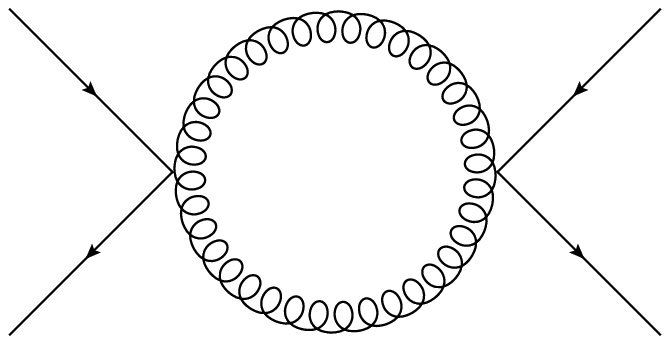}}
\put(2.67,6){\circle{0.5}}
\put(2.5,5.9){2a}
 \put(4,6){\includegraphics[scale=0.6]{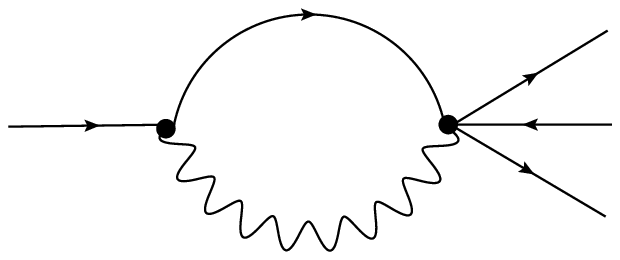}}
\put(6.87,6){\circle{0.5}}
\put(6.7,5.9){2b}
\put(0,3.5){\includegraphics[scale=0.5]{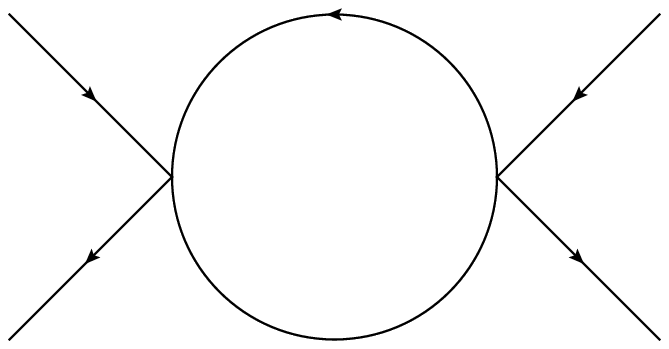}}
\put(2.67,3.5){\circle{0.5}}
\put(2.5,3.4){2c}
\put(4.5,3){ \includegraphics[scale=0.5]{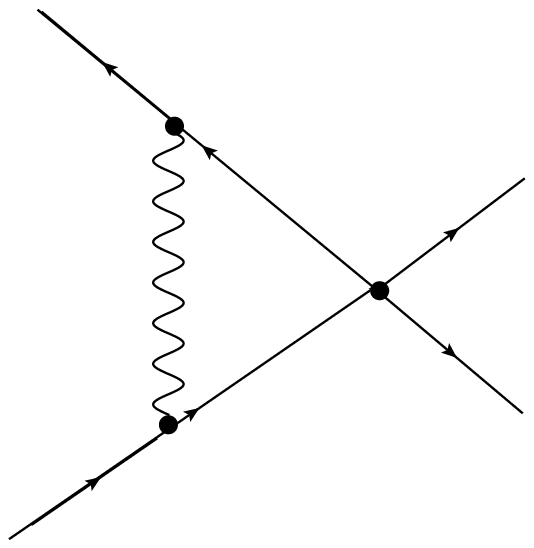}}
\put(6.87,3.5){\circle{0.5}}
\put(6.7,3.4){3a}
\put(0,1){\includegraphics[scale=0.5]{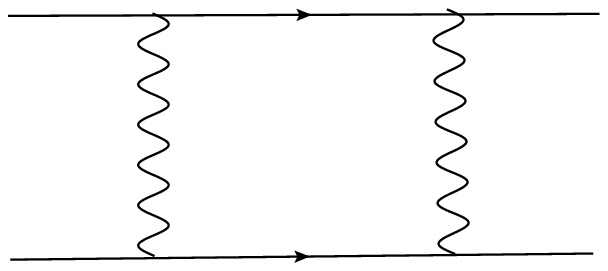}}
\put(2.67,0.7){\circle{0.5}}
\put(2.5,0.6){4a}
\put(4.5,1){\includegraphics[scale=0.5]{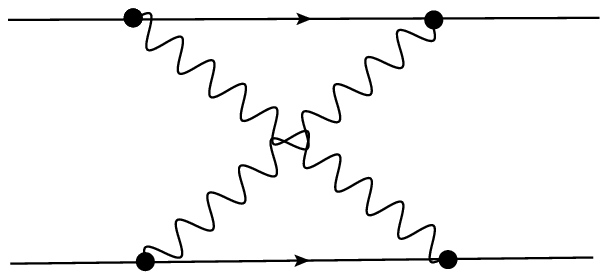}}
\put(6.87,0.7){\circle{0.5}}
\put(6.7,0.6){4b}
\end{picture}
\caption{Contributions to the running of the four-fermion couplings, sorted
  according to the number of vertices they contain. The diagrams containing
  curly lines receive contributions only from the trace mode, the diagrams
  with spiraling lines exist only for the TT mode.  The functional RG
  equation receives contributions from these diagram types with all internal
  lines and all vertices denoting full regularized propagators and vertices,
  respectively. The right-hand side of the flow is given by the $\tilde{\pat}$
  derivative of these diagrams, yielding corresponding regulator insertions in
  the internal propagators, c.f. \Eqref{eq:flowexp}.}
\label{diagrams}
\end{figure}

For the following discussion, we introduce the dimensionless renormalized
gravitational couplings
\begin{equation}
G=\frac{G_{\text{N}} k^2}{Z_{\text{N}}}
=  \frac{k^2}{32\pi\bar{\kappa}^2\,  Z_{\text{N}}},
\quad
\lambda= \frac{\bar{\lambda}}{k^2},
\label{eq:dimrencoup}
\end{equation}
and the corresponding anomalous dimension for the metric
\begin{equation}
\eta_{\mathrm{N}} = -\partial_t \ln Z_{\mathrm{N}}.
\label{eq:etas}
\end{equation}
Considering metric fluctuations within a general effective theory of quantum
gravity, these coupling constants can acquire a wide range of values near the
Planck scale. By contrast, these couplings are tightly constrained in the
asymptotic-safety scenario in the deep UV by the properties of a UV fixed
point. In fact, these fixed-point values are a result of the formalism and not
a parameter of the theory.  In this aspect, asymptotic safety is distinct from
many other approaches to quantum gravity, since here the microscopic action is
a prediction and not an assumption of the theory.  The anomalous dimension of
the background metric yields $\eta_N=-2$ as a necessary requirement for the
existence of the non-Gau\ss ian fixed point \cite{Reuter:1996cp}.  The fixed point values
$G_\ast,\lambda_\ast$ of the gravitational couplings depend on the
regularization scheme and also on the matter content. Within the
  regularization scheme and gauge choice used in \cite{Percacci:2002ie}, the
fixed-point values given in Tab.~\ref{tab:1} were obtained that 
    will be used in our analysis for illustration.

\begin{table}
\begin{tabular}{c|c|c|c|c|c|c|c}
$N_f$&1&2&3&4&5&10&25\\
\hline
$G_{\ast}$ &0.52&2.30&3.67&3.84&3.79&3.11&1.84\\
$\lambda_{\ast}$ &0.21&-1.38&-4.34&-6.18&-7.64&-12.33&-17.90
\end{tabular}
\caption{Fixed-point values for the gravitational couplings in the
  Einstein-Hilbert sector as a function of the number of chiral Dirac
  fermions, as computed in \cite{Percacci:2002ie}. }
\label{tab:1} 
\end{table}

In the following, we investigate the influence of metric fluctuations on the
chiral fermion sector taking both types of scenarios for quantum gravity into account.

\section{Flow of the fermion sector}\label{results}

Let us now discuss the flow in the fermion sector as induced by the various
diagrams sketched in Fig.~\ref{diagrams}. 

First we observe a cancelation between the two box diagrams (4a) and
(4b). This is not specific to gravity, but occurs whenever a Yukawa-type
fermion-antifermion-$\phi$ vertex with a scalar field $\phi$ without flavor (or
internal-symmetry) and/or Dirac indices, exists in the theory. As these box
diagrams involve only the trace mode, this cancelation mechanism between
ladder and crossed-ladder topologies is at work here.  A similar mechanism is
active in a non-chiral Yukawa coupling of the type $\phi \bar{\psi}^i
\psi^i$.

This marks an important difference to Yang-Mills interactions, as corresponding
box diagrams are the only contribution that generate the four-fermion
interaction even if they are set to zero initially. Since gravity gives rise to a
larger number of vertices from a minimally-coupled kinetic fermion
term, the diagram (2a) in figure \ref{diagrams}, being absent in Yang-Mills
theory, will create this interaction here.  Therefore the $\beta$ functions,
as given by \Eqref{betalambda} contain similar types of terms as in Yang-Mills
theory. 

Schematically, the $\beta$ functions for the dimensionful couplings
$\bar\lambda_\pm$ for a
general regulator $R_k(p^2)$ are given by
\begin{eqnarray}
 \beta_{\bar{\lambda}_\pm}&=& \partial_t \bar{\lambda}_{\pm}= 2 \Bigl[
\bar{\lambda}_{\pm} \frac{1}{32}
I[0,0,1]-\bar{\lambda}_{\pm}\frac{5}{8}I[0,1,0] \nonumber\\
&{}&\mp
\frac{15}{512 \cdot 8} I[0,2,0] -\bar{\lambda}_{\pm}
\frac{3}{16}I[1,0,1]\nonumber\\
&{}& +\bar{\lambda}_{\pm} 3
\frac{9}{256} I[2,0,1] + {\rm fermion\,\, loops}\Bigr].
\label{generalbetas}
\end{eqnarray}
The fermionic contribution indicated by the fermion-loop term corresponds to
the diagram class (2c) in Fig.~\ref{diagrams} and has first been calculated in
\cite{Gies:2003dp}.  The regulator-dependent dimensionful threshold functions
are defined by
\begin{eqnarray}
 I[n_f, n_{\rm TT}, n_{h}]&=& \tilde{\partial}_t \int
\frac{d^4p}{(2 \pi)^4}\,  (p^2)^n\, \frac{1}{\left(Z_{\psi}p^2 \left(1+r_k\left(\frac{p^2}{k^2}\right) \right)\right)^{n_f}}
\nonumber\\
&{}& \times\frac{1}{\left(\Gamma_{k\,
\text{TT}}^{(2)}\left(1+r_k\left(\frac{p^2}{k^2}\right) \right) \right)^{n_\mathrm{TT}}} \cdot \nonumber\\
&{}& \cdot\frac{1}{\left(\Gamma_{k\,
conf}^{(2)}\left(1+r_k\left(\frac{p^2}{k^2}\right) \right) \right)^{n_h}},\nonumber
\end{eqnarray}
where $n = n_\text{TT}+n_f+n_h-1$. In the above notation, we have
  already used regulators of the type $R_k=\Gamma^{(2)} r_k (y)$, with a
  dimensionless regulator shape function $r_k(y)$ and $y=p^2/k^2$.
Specializing to linear regulators of the type
\begin{eqnarray}
r_{k\, \rm grav}(p^2)&=& \left(\frac{\Gamma_k^{(2)}(k^2)}{\Gamma_k^{(2)}(p^2)}-1 \right) \theta (k^2 -p^2),\nonumber\\
r_{k\, \rm ferm}(p^2)&=& 
\left(\sqrt{\frac{k^2}{p^2}}-1 \right)\theta(k^2-p^2),
\end{eqnarray}
the fermionic flow equations for the dimensionless couplings $\lambda_\pm$,
can be determined explicitly
\begin{widetext}
\begin{eqnarray}
 \partial_t \lambda_{-}&=& 2 (1+\eta_{\psi})
\lambda_-+2 \Bigl[ -\frac{5G (\eta_N-6)}{24 \pi (1-2
\lambda)^2}\lambda_- - \frac{G (-6+\eta_N)}{4 \pi (3- 4
\lambda)^2}\lambda_- -\frac{5 G^2 (\eta_N-8)}{128
(-1+2\lambda)^3} \nonumber\\
&{}&+ \frac{G \left(36 \eta_N -7
\left(54-24 \lambda + \eta_{\psi}(-3+4 \lambda) \right)
\right)}{35 \pi (3-4 \lambda)^2}\lambda_- - \frac{9G
\left(21 \eta_N + 24 (-14+\eta_{\psi})-32
(-7+\eta_{\psi})\lambda \right)}{448 \pi (3-4
\lambda)^2}\lambda_-\Bigr]\nonumber\\
&{}&+ (-5+\eta_{\psi})
\frac{\lambda_-^2-{\rm N_f} \lambda_-^2-{\rm N_f}
\lambda_+^2}{40
\pi^2}
\end{eqnarray}
\begin{eqnarray}
 \partial_t \lambda_{+}&=& 2(1+ \eta_\psi)
\lambda_+ +2 \Bigl[-\frac{5G (\eta_N-6)}{24 \pi (1-2
\lambda)^2}\lambda_+- \frac{G (-6+\eta_N)}{4 \pi (3- 4
\lambda)^2}\lambda_+ +\frac{5 G^2 (\eta_N-8)}{128
(-1+2\lambda)^3}\nonumber\\
&{}&+ \frac{G \left(36 \eta_N -7
\left(54-24 \lambda + \eta_{\psi}(-3+4 \lambda) \right)
\right)}{35 \pi (3-4 \lambda)^2}\lambda_+- \frac{9G \left(21
\eta_N + 24 (-14+\eta_{\psi})-32 (-7+\eta_{\psi})\lambda
\right)}{448 \pi (3-4 \lambda)^2}\lambda_+\Bigr]\nonumber\\
&{}&+
(-5+\eta_{\psi})\frac{-2 \lambda_- \lambda_+ -2 N_f
\lambda_- \lambda_+ -3 \lambda_+^2}{40 \pi^2}.
\end{eqnarray}
\end{widetext}
Herein, the single terms correspond to the diagrams in fig.  \ref{diagrams} in
the following sequence: The first terms are the dimensional scaling terms of
$\lambda_{\pm}$. The first term in each square bracket corresponds to the
transverse traceless tadpole (1a), the second to the conformal tadpole
(1b). The third term in the square brackets that enters the two beta functions
with a different sign is represented by the two-vertex diagram (2a) with
internal metric propagators only. The mixed two-vertex diagram (2b) results in the
fourth term in square brackets. Finally the three-vertex diagram (3a)
corresponds to the last term in square brackets. The fermion-loop
contributions (2c) are represented in the two differing last terms; they agree
with \cite{Gies:2003dp}.

We find four pairs of (real) non-Gau\ss{}ian fixed points for $\lambda_{\pm}$ as a
function of $(G, \lambda, {\rm N_f},\eta_N,\eta_\psi)$. For gravity
approaching a non-Gau\ss ian fixed point, $G_{\ast} \neq 0$, the fermionic
Gau\ss{}ian fixed point is shifted and also becomes non-Gau\ss ian. If all
four fixed points persist also beyond this truncation, each one defines a UV
universality class of the fermionic matter sector. The fixed points can
quantitatively be classified by their number of relevant directions and the
corresponding critical exponents. 

The universal critical exponents can
be read off from the linearized form of the $\beta$ functions for general
couplings $g_i$ in the vicinity of the fixed point $g_{j\, \ast}$,
\begin{equation}
 \partial_t g_i = \sum_j
B_{ij}\left(g_j-g_{j\, \ast}\right) + \dots,\label{linflow}
\end{equation}
where the stability matrix $B_{ij}$is defined by
\begin{equation}
 B_{ij}= \frac{\partial
\beta_{g_i}}{\partial g_j} \Big|_{g= g_{\ast}}.
\end{equation}
\Eqref{linflow} is solved by
\begin{eqnarray}
g_i (k)= g_{i\,\ast}+ \sum_n C_n V_i^n \left( \frac{k}{k_0}
\right)^{-\theta_n}.
\end{eqnarray}
Herein the critical exponents $\{\theta\}=-
{\mathrm{spect}}(B_{ij})$ are minus the eigenvalues of the
stability matrix and $V^n$ are the (right) eigenvectors of
$B_{ij}$.
The scale $k_0$ is a reference scale and the $C_n$ are
constants of integration.

In order for the flow to hit a fixed point in the UV all $C_n$ pertaining to
\emph{irrelevant} directions with $\theta_n<0$ have to be set to zero. By
contrast, the $C_n$ for \emph{relevant} directions with $\theta_n>0$ are free
physical parameters that determine the long range physics.  As a consequence a
non-Gau\ss ian fixed point can be used to construct a predictive fundamental
theory if it has a finite number of relevant directions.

In the absence of gravity, the Gau\ss ian fixed point has two irrelevant
directions both with critical exponent
$\theta_{\text{Gau\ss}}=-2$,
corresponding to the standard powercounting canonical
dimension of the fermion
interactions in $d=4$ dimensions.
 The other three non-Gau\ss ian fixed points
all have at least one relevant direction with critical exponent $\theta=2$, as
can be proven on general grounds \cite{Gies:2003dp}. Two of these fixed points
have an additional irrelevant direction with negative critical exponent. The
last fixed point has another relevant direction with positive critical
exponent. Let us now discuss the effect exerted by metric fluctuations onto
the chiral fixed-point structure.

\subsection{Asymptotically safe quantum gravity}

 Let us analyze the fermionic flow in the asymptotic-safety scenario near the
  gravitational UV fixed point where $\eta_N=-2$ and $G$ and $\lambda$
  approach their fixed point values $G_\ast$, $\lambda_\ast$ as a function of
  the number of fermions $\Nf$ as determined in \cite{Percacci:2002ie},
  cf. Tab.~\ref{tab:1}. We also set $\eta_\psi=0$ here, as it is consistent
  with our truncation.

As one of our main results, the fermionic fixed-point structure persists under
the inclusion of metric fluctuations. Since the four-fermion couplings do not
couple back into the flow of the Einstein-Hilbert sector, the stability matrix
has a $2\times 2$ block of zeros off the diagonal.  Therefore the
gravitational and fermionic critical exponents are determined by the
eigenvalues in the Einstein-Hilbert sector and the fermionic subsector
separately.  Accordingly the gravitational critical exponents 
are given by the
well-known 
exponents
with positive real parts in the
Einstein-Hilbert sector 
(see \cite{reviews_AS} for an overview of typical
values without the effect of the minimally coupled fermions) and the two real critical exponents from the fermionic subsector. 

\begin{figure}[!here]
 \includegraphics[scale=0.8]{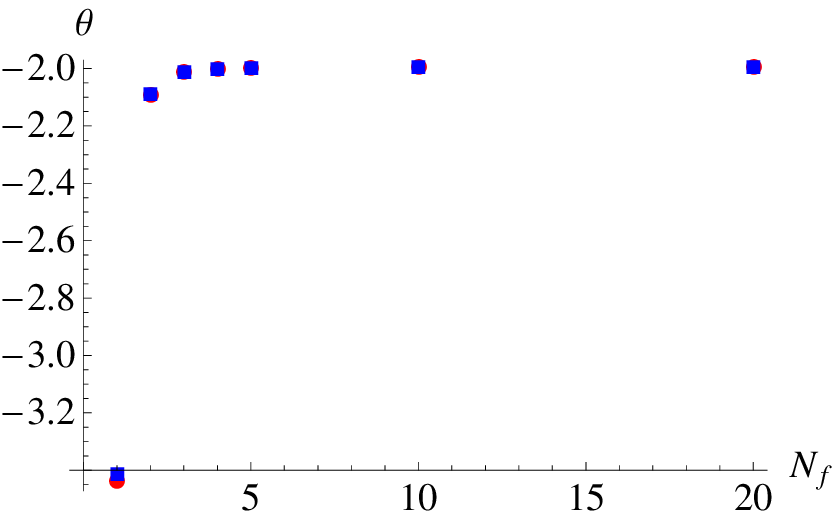}\newline\\
 \includegraphics[scale=0.8]{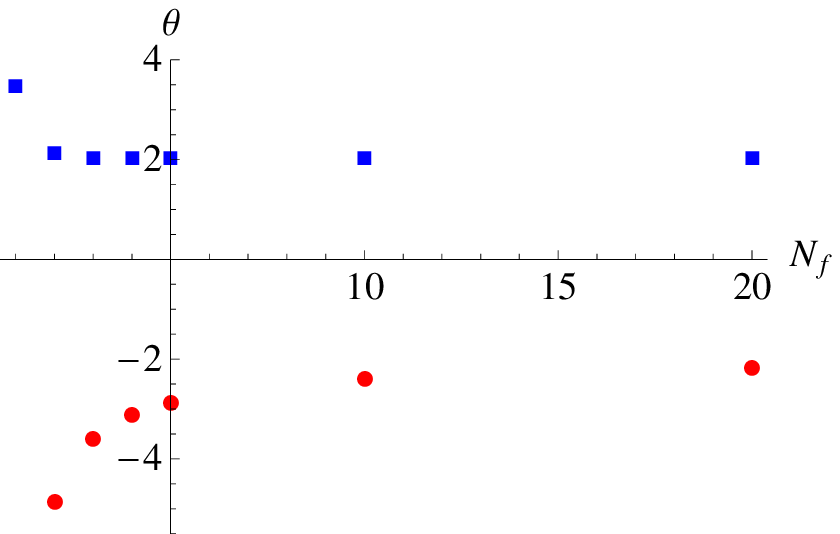}\newline\\
 \includegraphics[scale=0.8]{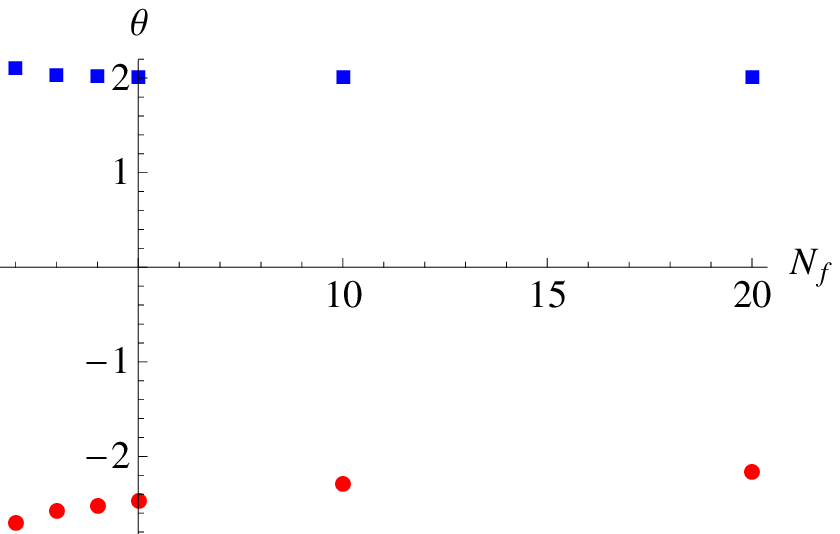}\newline\\
 \includegraphics[scale=0.8]{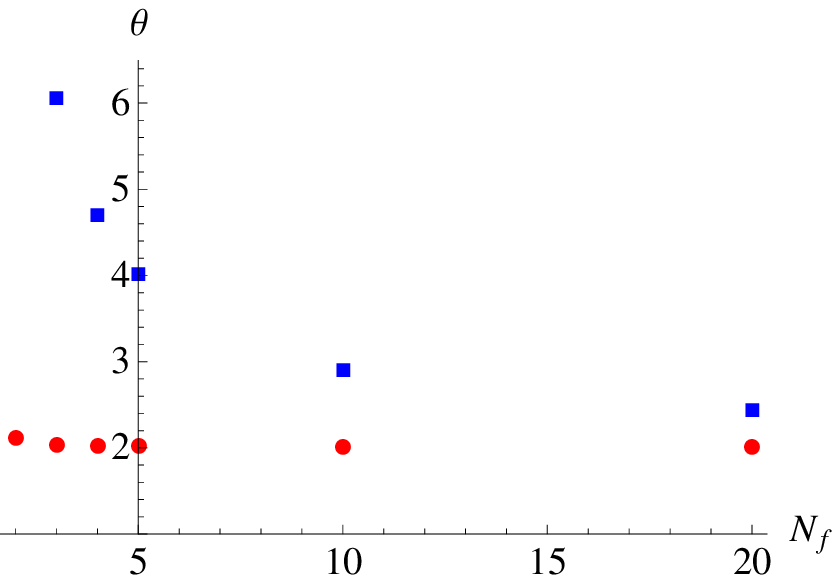}\newline
\caption{Critical exponents in the fermionic subsector as a
function of $\rm N_f$, where we always plot one critical
exponent with red dots and the second one with blue
squares. The upper plot corresponds to the shifted GFP and
therefore has irrelevant directions only.}
\label{fig:theta}
\end{figure}

Again we have four different fermionic fixed points at our disposal each
defining its own matter universality class, which have either two, one or no
relevant directions. The dependence of the critical exponents on $\rm N_f$ at
each of the four fixed points is shown in Fig.~\ref{fig:theta}.  These
critical exponents are determined by inserting the fixed point values of the
gravitational couplings taken from \cite{Percacci:2002ie} into the fermionic
part of the stability matrix. As these fixed point values are determined within a slightly different regularization scheme, this scheme-dependent error
adds to the systematic error of our truncation. Nevertheless, the general
chiral fixed point structure is rather insensitive to variations of the
non-universal input. 

The critical exponents approach the limiting values of the purely fermionic
system for large $N_f$. This is due to the following mechanism being at work
here: As shown in \cite{Percacci:2002ie}, the backreaction of a minimally
coupled fermion sector onto the Einstein-Hilbert sector shifts $\lambda_\ast$
to increasingly negative values as a function of $N_f$. In the propagators, a
negative value for $\lambda$ acts similarly to a mass term for the
metric. This suppresses the contribution from metric fluctuations to
$\beta_{\lambda_{\pm}}$ for large $\Nf$. 
This decoupling mechanism induced by
an increasingly negative cosmological constant ensures that the properties of the matter
sector will not be strongly altered by metric fluctuations.

\begin{figure}[!here]
 \includegraphics[scale=0.7]{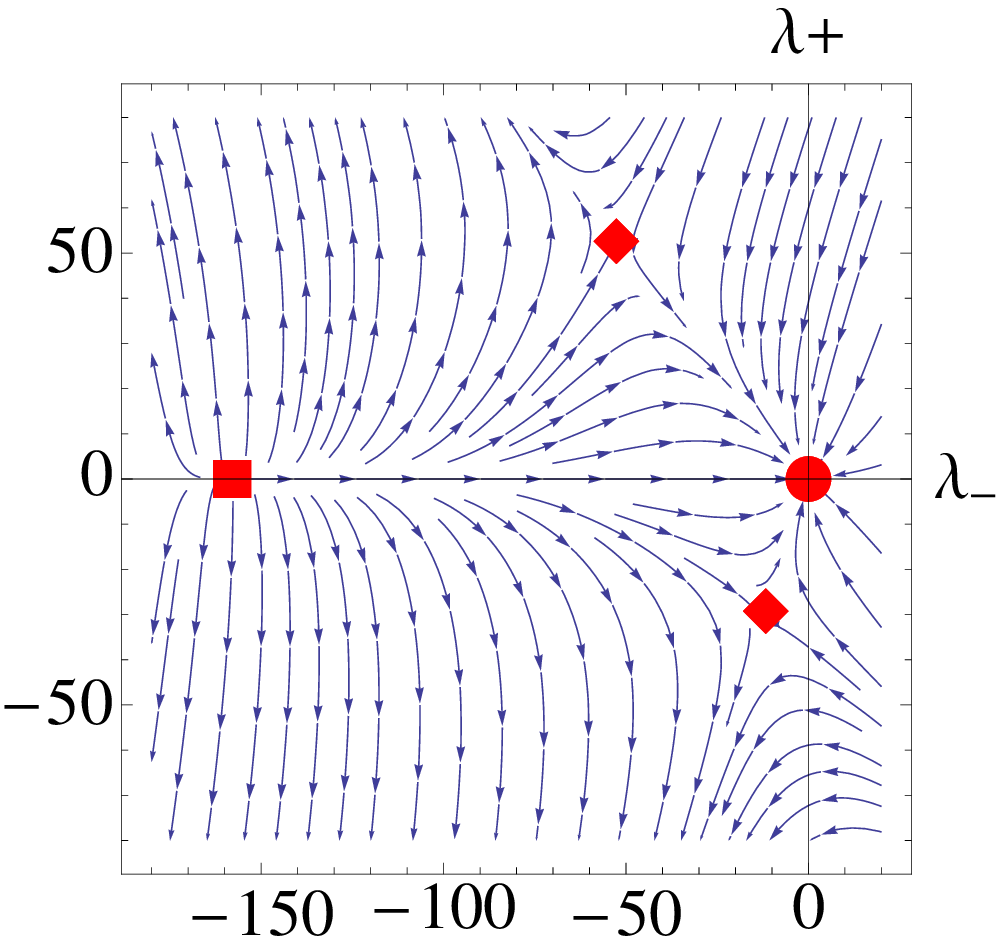}
 \includegraphics[scale=0.7]{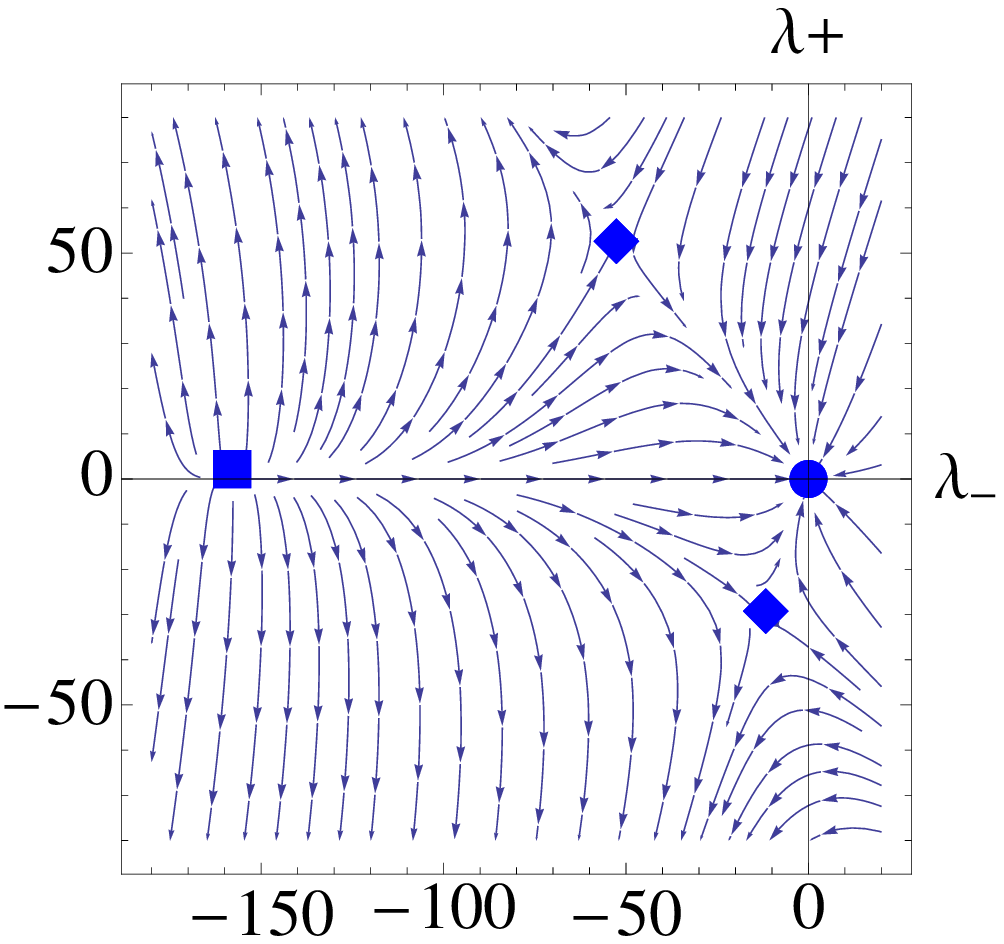}
\caption{Flow lines towards the infrared in the chiral $(\lambda_+,
\lambda_-)$ plane for $\Nf=2$. The upper panel shows the flow with
$G=0=\lambda$, the lower one with $G\approx 2.3$ and
$\lambda \approx -1.38$. Both panels differ only slightly,
since the decoupling mechanism due to the negative cosmological
constant induces a strong suppression of
the metric fluctuations. Dots represent the shifted Gau\ss ian fixed point
with two irrelevant directions, diamonds/squares depict non-Gau\ss ian fixed points
with one/two relevant directions.}
\label{flows}
\end{figure}

This chiral fixed-point structure is illustrated in Fig.~\ref{flows}, 
where the fixed point positions and the RG flows towards the infrared are
depicted in the $(\lambda_+, \lambda_-)$ plane for $N_f=2$.
The pure fermionic flow in the upper panel differs very little from the
corresponding flow including the metric fluctuations in the gravitational
fixed point regime (lower panel). Apart from minor shifts of the fixed point
positions, the flow diagrams are very similar. 

It should be emphasized that the decoupling mechanism due to a negative
cosmological constant is only active in theories with a dominant number of
fermionic degrees of freedom, as it is the case for the standard model.
Since minimally coupled scalars shift the fixed-point
value for the cosmological constant towards $\lambda_{\ast}\rightarrow
\frac{1}{2}$ \cite{Percacci:2002ie}, a larger number of scalars even results in an enhancement of metric fluctuations.
As a consequence, even at the shifted
Gau\ss{}ian fixed point the fermionic system can develop strong correlations,
since the fixed point values for $\lambda_{\pm}$ can then become quite large
(cf Fig.~\ref{lambdaplusplot} below).
Accordingly, in theories with a supersymmetric
matter content and low-scale supersymmetry breaking, such a decoupling
mechanism might not occur or only in a much weaker fashion. Supersymmetric theories with a quantum
gravity embedding may thus have to satisfy stronger constraints as far as the
initial conditions of their RG flow are concerned.

\subsection{General effective quantum gravity theories}

Let us broaden our viewpoint a bit by considering a wider range of effective
theories of quantum gravity. Consider an unknown UV completion of quantum
gravity, which may differ strongly from a description in a QFT framework,
e.g., by introducing a physical discreteness scale for spacetime, possibly
also including the matter sector in a unified manner. Still, effective metric
degrees of freedom may be expected to become relevant at a scale $k_0$ close
to or below the Planck scale (see also \cite{Percacci:2010af}). The
underlying microscopic theory may effectively reduce to a quantum field theory
for matter and effective metric degrees of freedom with gravity rapidly
becoming semi-classical towards the IR. Without any knowledge about the
underlying theory, the effective quantum field theory may be at any point in
theory space.

For scales $k<k_0$ accordingly our present framework becomes applicable even
for these different approaches to quantum gravity. Within this 
framework, we thus allow for any value of the chiral couplings and admissible
values of the gravitational couplings (i.e., a positive Newton constant, and
$\lambda< 1/2$ in order to avoid the potentially artificial propagator
poles). The anomalous dimensions $\eta_N$ and $\eta_\psi$, representing {\em
  inessential} couplings, will be determined by the effective field
theory. Nevertheless, in our truncated framework, we allow them to acquire
a priori unknown values of $\mathcal{O}(1)$.

Following the RG flow further towards the IR, the fermionic sector should
approach the Gau\ss{}ian fixed point in standard scenarios in order to
be weakly correlated on scales where metric fluctuations become unimportant,
and gauge boson and matter fluctuations start to dominate the picture. This
requires that the initial values of the four-fermion couplings, which can (in
principle) be determined from the underlying microscopic theory, have to lie
within the basin of attraction of the Gau\ss{}ian fixed point. This presents a
non-trivial requirement that has to be fulfilled by any theory of quantum
gravity.

\begin{figure}[!here]
 \includegraphics[scale=0.7]{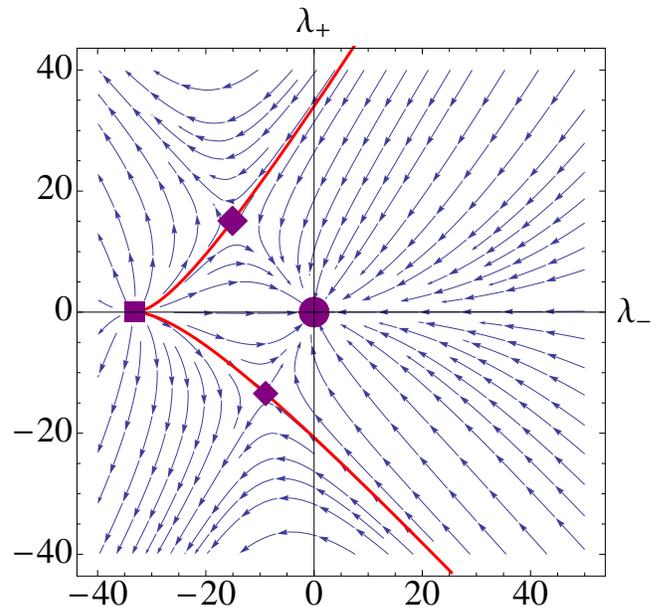}
\caption{Flow towards the infrared in the $\lambda_{+},
  \lambda_{-}$-plane for $\eta_N =0, \eta_{\psi}=0, G=0.1, \lambda=0.1$ and
  $N_f=6$. For initial values to the right of the red lines the chiral system
  is in the universality class of the (shifted) Gau\ss ian fixed point. Any
  microscopic theory that would put the effective quantum field theory to the
  left of the red lines would generically not support light fermions.}
\label{fig:basin}
\end{figure}

As a specific example, we depict the basin of attraction for $\eta_N=0,
\eta_{\psi}=0, N_f =6, G=0.1$ and $\lambda=0.1$ in Fig.~\ref{fig:basin}

\begin{figure}[!here]
\includegraphics[width=0.5\textwidth]{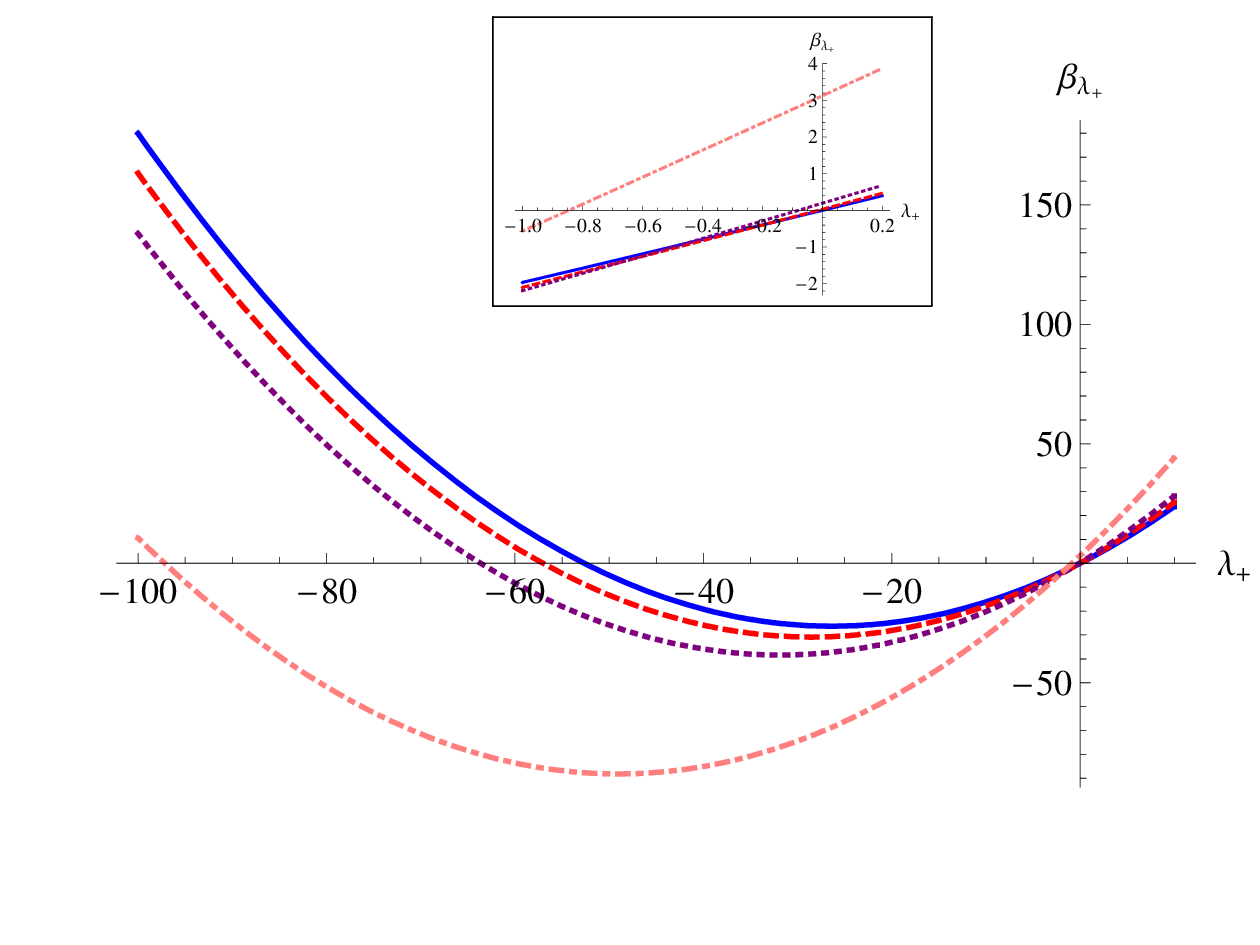}
\caption{$\beta_{\lambda_{+}}$ as a function of $\lambda_{+}$ for different
  values of $G$. The various curves correspond to $G=0$ (full blue), $G=0.2$
  (red dashed), $G=0.5$ (purple dotted), and $G=2$ (dotdashed). The inlay
  shows the region around the Gau\ss{}ian fixed point, illustrating a
  significant shift from the non-interacting limit.}
\label{betalambda}
\end{figure}

For generic values of $G$ and $\lambda$ the system can be altered
considerably.  As a first observation, the parabolas characterizing the
fermionic $\beta$ functions broaden as a function of
$G$ for a fixed value of $\lambda$, as shown in Fig.~\ref{betalambda}.  In
this example, we plot the $\beta$ function for $\lambda_{+}$ for fixed
$N_f=2$, $\eta_N=-2$, $\eta_{\psi}=0$ and $\lambda=0$. We set $\lambda_{-}$ on
the shifted Gau\ss{}ian fixed point value.

A particularly strong effect can be observed for positive values of $\lambda$.
Here, the contribution from the metric sector is further enhanced for $\lambda
>0$. Indeed the $\beta$ functions show the well-known divergence for
$\lambda = \frac{1}{2}$. Whereas the divergence is likely to be an artifact of
the simple Einstein-Hilbert truncation in the IR, the response of the
gravitational propagators naturally suggests an enhancement of metric
fluctuations for positive cosmological constant $\lambda$.

\begin{figure}[!here]
 \includegraphics[scale=0.7]{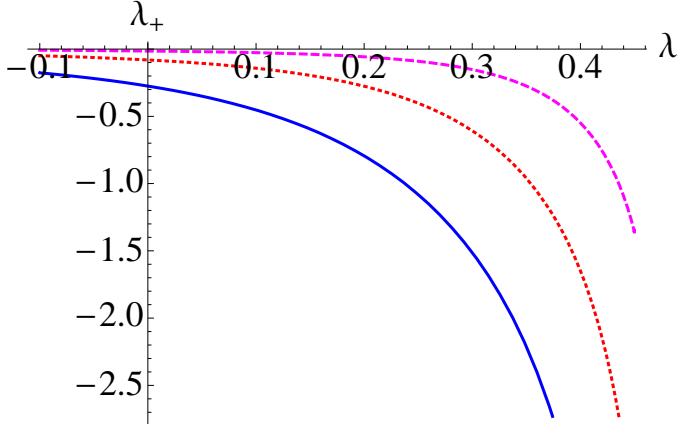}
\caption{Fixed-point value of $\lambda_+$ at the shifted Gau\ss ian fixed
  point for $N_f=2$, $\eta_{\psi}=0$ and $\eta_N=0$ as a function of $\lambda$
  for $G=1$ (full blue line), $G=0.5$ (red dotted line) and $G=0.2$ (magenta
  dashed line).}
\label{lambdaplusplot}
\end{figure}

As an example, we plot the value of $\lambda_+$ at the
shifted Gau\ss{}ian fixed point in Fig.~\ref{lambdaplusplot}.

\begin{figure}[t]
\includegraphics[scale=0.7]{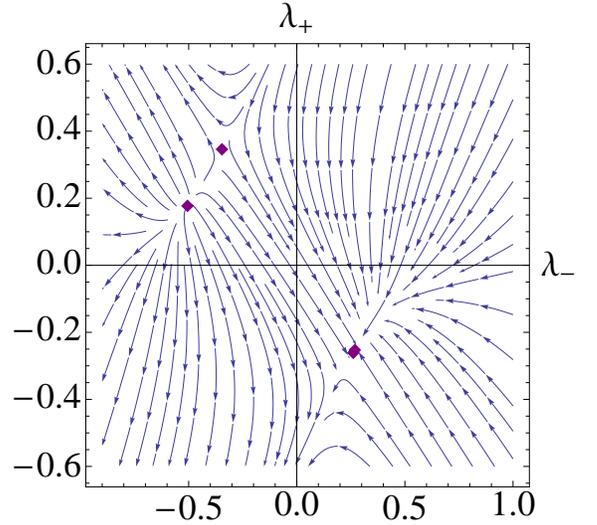}
\caption{At $\eta_{\psi} \approx -1.0592$ (for the specific parameter
    values $N_f=6, G=0.1, \lambda=0.1, \eta_N =-2$) the Gau\ss{}ian fixed
  point and a fixed point with one relevant direction fall on top of each
  other (in the lower right quadrant).}
\label{flow_neg_eta} 
\end{figure}

We observe that the Gau\ss{}ian fixed point can be shifted to considerably
larger values of $(\lambda_-, \lambda_+)$.  This implies that the system will
be strongly-interacting in this sector even at the fixed point which
corresponds to the Gau\ss ian fixed point in the absence of gravity. This may
exert a strong influence on the physics properties of this shifted Gau\ss{}ian
universality class.  As an example for such a significant deformation, we find
that a negative anomalous dimension can in principle induce a crucial change
in the fixed point structure. In particular the shifted Gau\ss{}ian fixed
point "collides" with a fixed point with one relevant direction at a negative
value for $\eta_{\psi\, \rm crit}$ (see Fig.~\ref{flow_neg_eta}). At
  this point, one critical exponent of each fixed-point approaches zero.

For values $\eta_{\psi}< \eta_{\psi\, \rm crit}$ the two fixed points move
apart again. Assuming that in the regime of strong gravity fluctuations these
induce a large negative value for $\eta_{\psi}$, the flow towards the
semiclassical infrared would
then cross $\eta_{\psi \, \rm crit}$. Accordingly points that lie in what at a
first glance seems to be the basin of attraction for the shifted Gau\ss{}ian
fixed point can leave this region during the flow. Precise constraints
  for such quantum gravity theories and the associated chiral sector then
  depend on the dynamical details of the full flow.

\subsection{Metric vs. gauge boson fluctuations}

At first sight, the absence of gravity-stimulated chiral symmetry breaking
seems surprising. Since gravity leads to attractive forces between matter, it
is plausible to expect that binding phenomena are enhanced upon the inclusion
of gravity. It is therefore instructive to confront our results with the
chiral symmetry breaking mechanism in QCD-like theories. 

Several technical differences have already become obvious: (i) the fermion
anomalous dimension $\eta_\psi$ does not vanish in gravity in the
Landau-DeWitt gauge (even though we have set it to zero in
the present
approximation).
(ii) ladder and crossed-ladder box diagrams ((4a) and (4b) in
Fig.~\ref{diagrams}) cancel in gravity, but play an important role in the
approach to chiral symmetry breaking in QCD.

Further differences can be read off from the above analysis. For the
fixed-point annihilation to occur (cf. Fig.~\ref{fig:parabolasketch}) the
terms $\sim G^2$ in the fermionic flows ((2a) in Fig.~\ref{diagrams}) have to
dominate. In the case of gravity, they are however outweighed by the tadpole
terms $\sim G \lambda_\pm$.

In more physical terms, the $\sim G^2$ terms describe the attractive nature of
gravity, whereas the tadpole terms $\sim G\lambda_\pm$ play the role of a
gravity contribution to the (anomalous) scaling of the fermion couplings,
$\pat \lambda_\pm = + 2(1+\eta_\psi+ \dots G)\lambda_\pm+ \dots$. Quite
generally, the fixed point structure in the fermionic flows for $d>2$ arises
from a balancing between dimensional scaling $\pat\lambda_\pm \sim
\lambda_\pm$ and fermion fluctuations $\sim \lambda_\pm^2$. Whereas
gauge-field fluctuations support the fermionic fluctuation channels, metric
fluctuations also take a strong influence on the anomalous dimensional scaling
which counteracts the general attractive effect of gravity.

This viewpoint is further supported by other technical observations: whereas
gravity is channel blind with respect to the scaling terms, i.e., $\pat
\lambda_i \sim G\lambda_i$, gauge boson fluctuations with coupling $g$ also
give rise to terms $\pat \lambda_i \sim g^2 \lambda_j$ with $i\neq j$ that
rather act like the above mentioned fluctuation terms. Finally, we should
mention that there are further examples that fluctuations of attractive forces
do not necessarily support binding phenomena: e.g., an effective flavored
chiral Yukawa interaction in QCD-like theories contributes via box diagrams
with a sign opposite to that of gauge bosons to the fermionic flow
\cite{Gies:2002hq}.

Still, it should be kept in mind that the present analysis is carried out in a
restricted truncation of the effective action that -- though meaningful for
QCD-like theories --  might not be sufficient for gravity. Further operators
which could potentially be relevant for the interplay between gravity and a
chiral fermion sector are discussed below. 

\section{Conclusions}\label{conclusions}

We have investigated the quantum interplay between chiral fermions and metric
fluctuations in quantum gravity. In contrast to QCD-like systems, where gluon
fluctuations can induce strong fermionic correlations leading to chiral
symmetry breaking, metric fluctuations do not support this mechanism in an
analogous framework. Our result thus indicates that the existence
and observations of light fermions is well compatible with a regime, e.g.,
near or above the Planck scale, where quantum gravity effects in the form of
sizeable metric fluctuations set in. 

More specifically, light fermions are compatible with the asymptotic-safety
scenario of quantum gravity which provides a UV completion of quantum gravity
within quantum field theory. In particular, we observe a decoupling mechanism for metric fluctuations:
As known from \cite{Percacci:2002ie}, the
dominant effect of fermionic fluctuations is a shift of the gravitational
non-Gau\ss ian fixed point towards increasingly negative values of the
dimensionless renormalized cosmological constant for larger numbers of flavors
$\Nf$. As a consequence, metric fluctuations decouple from the matter
sector as such a cosmological constant acts like a mass term in the
propagators of the metric modes and thus suppresses metric fluctuations. Apart from a slight shift of the Gau\ss ian
matter fixed point to small but non-vanishing values of the fermion
interactions, the universality class of the Gau\ss ian fixed point which is
supposed to describe the fermionic matter content of the universe is left
rather unaffected. In particular the critical exponents remain very close to
the Gau\ss ian values. 

Such a compatibility scenario between matter and asymptotically safe gravity
holds at least as long as fermions remain the dominant degrees of freedom. For
larger numbers of bosonic modes, in particular for a supersymmetric matter
content, the whole system may behave differently. 

Our results are also applicable to scenarios where quantized metric
fluctuations are considered as part of an effective quantum field theory at an
intermediate scale below or near the Planck scale.  Here the underlying
  UV completion of gravity, which may use different degrees of freedom for
  gravity than the metric, or even leave the local QFT framework completely,
  will determine the initial conditions for the RG flow at this intermediate
  scale.  Also from this more general viewpoint, we do not find any
indications for gravity-stimulated chiral symmetry breaking. Still, our
results can be used to decompose the accessible theory space into those
branches where the chiral sector remains symmetric and other branches where
the chiral sector becomes critical and typically generates heavy fermion
masses. The distribution of these branches in theory space depends on the
gravitational couplings. In particular, the universality properties of the
shifted Gau\ss ian matter fixed point can substantially vary. This analysis
provides general constraints on the Planck scale behavior of any microscopic
theory of quantum gravity: the existence and observation of light fermions
potentially excludes those branches of theory space where the chiral sector is
critical at the Planck scale.

Of course, a variety of further aspects could modify our results
quantitatively and qualitatively, as our analysis is performed in a limited
and comparatively small hypersurface in theory space. Still, the interplay
between sizeable metric fluctuations and chiral symmetry has the potential to
provide relevant phenomenological constraints for any theory of quantum
gravity also in a more complete and quantitatively reliable investigation. Let
us try to list some issues that need to be addressed in further studies:

Whereas we have used flat-space calculations mainly as a technical tool, the
full theory of quantum gravity will predict an (effective) manifold as its
solution to the equations of motion. This resulting background may take
influence on the chiral status of the matter sector itself, as it may screen
or enhance fermionic long-range fluctuations that lead to chiral
criticality. Screening mechanisms for chiral symmetry breaking of this type
have already been studied in various chiral models, such as two- and
three-dimensional (gauged) Thirring models
\cite{Sachs:1993ss,Sachs:1995dm,Geyer:1996yf} or the four-dimensional gauged
NJL model \cite{Geyer:1996kg}. Depending on its sign, curvature can
act like an IR cutoff that screens the critical IR fluctuations. 

Also, if the question of chiral symmetry breaking and restoration is
considered in a cosmological context, the thermal evolution of the universe
may play an essential role. Broken chiral symmetry could be restored during
reheating, if the corresponding temperature is sufficiently high compared with
the scale of critical fermion dynamics -- independently of whether it is
stimulated by metric fluctuations or another mechanism.

Let us also discuss the possibility of gravity-stimulated chiral symmetry
breaking which would strongly differ from the scenario arising from QCD-like
theories. As gravity supports a richer structure of operators, chiral
criticality could be triggered by operators that are typical for gravity, but
do not occur for other theories. 

Up to canonical dimension four and five, only explicitly symmetry breaking
terms (as e.g. $\int d^4x \,\sqrt{g}\, R \bar{\psi}^i \psi^i$)
exist. 

At dimension six (for two-fermion terms) and eight (for four-fermion-terms) we encounter a variety of new terms that
are not forbidden by explicit chiral symmetry breaking,
for instance
\begin{eqnarray}
\text{dim 6:} &\quad&\int d^4x \, \sqrt{g} \,R\, \bar{\psi} \slashed{\nabla}\psi,\nonumber\\
 &{}&\int d^4x \, \sqrt{g}\, R_{\mu \nu}\, \bar{\psi}
\gamma^{\mu}\nabla^{\nu}\psi, \label{extended_trunc}
\\
\text{dim 8:}&\quad&\int d^4x \, \sqrt{g} \, R \left(V^2 \pm A^2 \right),\nonumber\\
&{}&\int d^4x \, \sqrt{g}\, R_{\mu \nu} \Bigl(\left( \bar{\psi}^i
\gamma^{\mu}\psi^i \right)\left( \bar{\psi}^j \gamma^{\nu}\psi^j \right) \nonumber\\
&{}& \phantom{xxxxx} - \left( \bar{\psi}^i
\gamma^{\mu}\gamma^5\psi^i \right)\left( \bar{\psi}^j
\gamma^{\nu}\gamma^5\psi^j \right)\Bigr).
\label{extended_trunc2}
\end{eqnarray}
At higher dimensionalities the number of terms increases considerably, as then
also e.g. contractions involving the Riemann tensor will be
possible. Furthermore couplings involving $\slashed{\nabla}$ or higher
  powers of the curvature are possible.  Distinguishing between the
background and the fluctuation metric leads to an even larger "zoo" of
possible operators.

\begin{figure}[t]
\setlength{\unitlength}{1cm}
\begin{picture}(8,5)
\put(2.5,3){\includegraphics[scale=0.5]{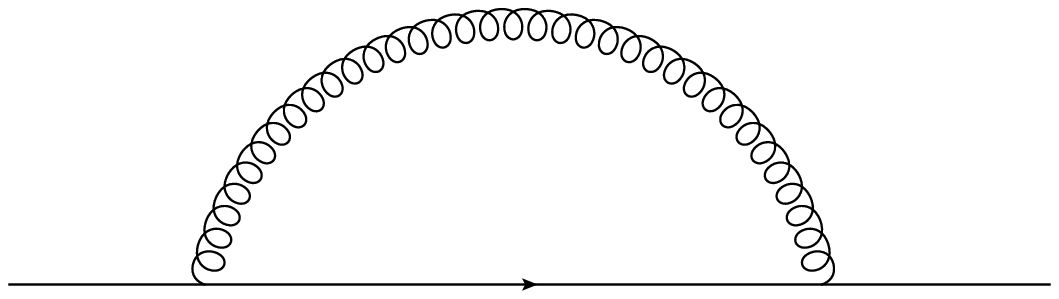}}
\put(0.2,3.5){dim 6:} 
\put(2.5,0){\includegraphics[scale=0.5]{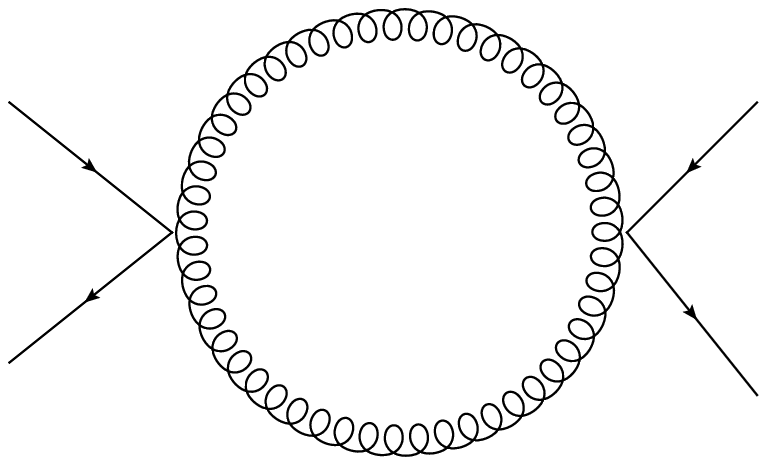}}  
\put(0.2,1){dim 8:}
\end{picture}
\caption{2-fermion and 4-fermion couplings are generated from metric
  fluctuations. Here we have not drawn external graviton lines/couplings to a
  non-trivial background curvature; these will be generated by taking
  derivatives of the above diagrams with respect to the desired metric
  structure. The results then correspond to dimension 6 and 8 operators
    of the type listed in Eqs.~\eqref{extended_trunc} and
    \eqref{extended_trunc2}. The upper self-energy diagram on flat space also
    contributes to the fermion anomalous dimension $\eta_\psi$. Similar
  diagrams occur at higher order in the fermion field.}
\label{fig:8}
\end{figure}

Several comments are in order here: The effect of metric
fluctuations implies that none of these couplings will have a
Gau\ss{}ian fixed point, as the
antifermion-fermion-two-graviton vertex generically
generates these couplings even if they are set to zero. The corresponding
diagrams are indicated in Fig.~\ref{fig:8}.

These couplings raise several issues: In order for the 
asymptotic-safety scenario to work, the complete system of
gravitational couplings, four-fermion-couplings and mixed
couplings such as the above ones has to admit a
Non-Gau\ss{}ian fixed point (even though this does not
necessarily require $g_{i\, \ast}\neq 0$ for all possible
couplings). As the above couplings couple non-trivially into
the flow of the Einstein-Hilbert action and the four-fermion
couplings, they may change our findings in this sector.

Since in the truncation that we have studied here, chiral symmetry breaking is
avoided as the gravitational contribution to the anomalous scaling of the
fermionic couplings outweighs the contribution that triggers bound-state
formation, we expect that in particular the dimension-6 non-minimal kinetic
terms in \Eqref{extended_trunc} represent an interestion extension of our
truncation specific to gravity. This is because these generate further
contributions $\sim G^2$ and do not contribute to the anomalous scaling.  As
we expect that these couplings are non-zero at any UV fixed point, they
constitute a non-vanishing contribution to the $\beta$ functions of the
four-fermion couplings. In other words, these dimension-6 terms genuine to
gravity have the potential to act structurally identical in the fermionic
flows, as the fermionic self-interaction terms considered so far.

Another question that has remained unaddressed so far, is the question of
gravity-induced symmetry-breaking patterns. In the present work, we have
imposed a rather standard chiral SU$({\rm N_f})_{\mathrm{L}}$ $\times$
SU$({\rm N_f})_{\mathrm{R}}$ (with additional U(1) factors of particle number
and axial symmetry), and implicitly assumed its breaking in a QCD-like
fashion, i.e. to a remaining mesonic SU($\Nf$) symmetry. Other breaking
patterns are conceivable, including an originally larger symmetry that may
break to the standard chiral symmetry upon large metric fluctuations. Our work
merely represents a first step in this direction. In principle, it seems
worthwhile to think not only about gravitationally-stimulated symmetry
breaking and corresponding condensates, but also about corresponding bound
states or excitations on top of condensates. If a gravitationally-stimulated
symmetry breaking transition with a remnant standard chiral symmetry occurred
near the Planck scale, stable bound states (analogously to hadrons in QCD) may
have remained and (if equipped with the right quantum numbers) could
contribute to the dark matter in the universe.  

Furthermore, in analogy to recent ideas in QCD, where a quarkyonic phase with
confinement but intact chiral symmetry supports a spectrum of bound states,
bound states may form that correspond to bosonized operators, e.g. of the
form \Eqref{extended_trunc2}. These might form at a scale where quantum
gravity is strongly interacting, and may then become massive at the much lower
scale of chiral symmetry breaking. Supporting a stable bound state over such
a large range of scales requires, of course, a highly non-trivial
interplay between gravity and matter.

\acknowledgments

Helpful discussions with Michael M. Scherer, Jens Braun, and Andreas Wipf
are gratefully acknowledged. This work was supported
by the DFG-Research Training Group "Quantum- and
Gravitational Fields" (GRK 1523/1) and by the DFG grant Gi 328/5-1 (Heisenberg
program), and the DFG research unit FOR 723.

\begin{appendix}\label{appendix}
 \section{Variation of the effective action}
We expand the vierbein around a (flat) background:
\begin{equation}
 e_{\mu a}= \bare_{\mu a}+ \delta e_{\mu a},\label{vierbeinfluc}
\end{equation}
where higher orders are not needed in our calculation.
In the following we choose the Lorentz symmetric gauge with gauge-fixing functional \cite{Woodard:1984sj}\cite{vanNieuwenhuizen:1981uf}, as then all vierbein fluctuations can be rewritten in terms of metric fluctuations without ghosts due to the $O(4)$ gauge fixing:
\begin{equation}
 F_{ab}= e_{\mu a}\bar{g}^{\mu \nu}\bare_{\nu b}- e^{\mu b}\bar{g}^{\mu \nu}\bare_{\nu a} \label{vierbeingauge}.
\end{equation}
This allows to write
\begin{eqnarray}
\delta e_{\mu a}&=& \frac{1}{2}h_{\mu}^{\,\,\kappa}\bar{e}_{\kappa a}\label{deltavar}\\
\delta e^{\kappa b}&=& -\frac{1}{2} h_{\mu}^{\,\,
\kappa}\bar{e}^{\mu b}\label{deltainvvar}\\
\end{eqnarray}
We also have that
\begin{eqnarray}
\phantom{,}[\gamma^a,\gamma^b]\, 
\delta \omega_{\mu ab}&=& [\gamma^{\lambda}, \gamma^{\nu}]D_{\nu}h_{\lambda \mu}\label{omegavar}.
\end{eqnarray}

From \eqref{omegavar} we can deduce for constant external fermions, where
total derivatives can be discarded, that
\begin{eqnarray}
\lbrack\gamma^a, \gamma^b\rbrack\delta^2 \omega_{\mu ab}&=& [\gamma^{\lambda}, \gamma^{\nu}]\Bigl(-h^{\sigma}_{\, \lambda}D_{\nu}h_{\mu \sigma}-h^{\sigma}_{\, \nu}D_{\sigma}h_{\mu \lambda}\nonumber\\
&{}&-\frac{1}{2}h_{\kappa \lambda}D_{\mu}h^{\kappa}_{\, \nu} \Bigr),\label{omega2var}
\end{eqnarray}
where we have set $g_{\mu \nu}= \bar{g}_{\mu \nu}$ and $e_{\mu a}= \bare_{\mu
  a}$, and then dropped the bar on the covariant derivative.

We then go over to Fourier space
\begin{eqnarray}
 \psi(x)&=& \int \frac{d^4p}{(2 \pi)^4}\psi(p)e^{-ipx}\nonumber\\
 h_{\mu \nu}(x)&=& \int \frac{d^4p}{(2 \pi)^4}h_{\mu \nu}(p)e^{-ipx}\nonumber\\
 \bar{\psi}(x)&=& \int \frac{d^4p}{(2 \pi)^4}\bar{\psi}(p)e^{ipx},\nonumber\\
\end{eqnarray}
where $\psi(x)$ and $\psi(p)$ denote Fourier transforms of each other.

Now we may evaluate the mixed fermion-graviton
vertices, where our conventions are
\begin{eqnarray}
\Gamma^{2}= \frac{\overset{\rightarrow}{\delta}}{\delta
\Phi^T(-p)}\Gamma \frac{\overset{\leftarrow}\delta}{\delta
\Phi(q)},
\end{eqnarray}
where the collective fields
\begin{eqnarray}
 \Phi^T(-q)&=& \left(h_{\kappa \lambda}^{\mathrm{TT}}(-q), h(-q),
 \psi^T_i(-q), \bar{\psi}_i(q) \right)\\ 
\Phi(q)&=&\left(h_{\mu \nu}^{\mathrm{TT}}(q), h(q),\psi_j(q),
\bar{\psi}^T_j(-q) \right). 
\end{eqnarray}
Here the second line should be read as a column vector. The symbol $T$ refers
to transposition in Dirac space and in field space. 
As we work in the Landau deWitt gauge, only the transverse
traceless and the trace mode can contribute.

The first variation of the kinetic fermion term with respect to the metric is given by
\begin{eqnarray}
 \delta \Gamma_{\rm kin}&=& i Z_{\psi}\int d^4 x \bar{\psi}^i\Bigl(\delta(\sqrt{g}) \gamma^{\mu}\nabla_{\mu}+ \sqrt{g}\delta \gamma^{\mu}\nabla_{\mu}\nonumber\\
&{}&+ \sqrt{g}\gamma^{\mu}\delta \nabla_{\mu} \Bigr)\psi^i. \label{varkin}
\end{eqnarray}
To read off the trace-mode-fermion-vertices we
Fourier-transform the first variation of the kinetic term
with respect to the metric to get (in agreement with
\cite{Zanusso:2009bs})

\begin{eqnarray}
 \delta \Gamma_{\rm kin}&=& Z_{\psi} \int
\frac{d^4p}{(2\pi)^4}
 \Bigl(\Bigl(\frac{3}{16}h(p)\bar{\psi}^i(p)\slashed{p}
\psi^i\nonumber\\
&{}&
-\frac{3}{16}\bar{\psi}^i\slashed{p}
\psi^i(-p)h(p)\Bigr)\Bigr).\label{firstvar}
\end{eqnarray}
In this notation, $\bar{\psi}$ and $\psi$ are the constant background fields, whereas the momentum-dependent fluctuation fields are distinguished by carrying an appropriate argument.
This allows to evaluate the following vertices:
\begin{eqnarray}
 V_{\rm kin}^{h\, \bar{\psi}^{i\, T}}&=&
\frac{\delta}{\delta h(-p)}
\,\Gamma_{\rm kin}\frac{\overset{\leftarrow}{\delta}}{\delta
\bar{\psi}^{i\, T}(-q)}= \frac{3}{16}Z_{\psi}\psi^{i\,
T}\slashed p^T\\
V_{\rm kin}^{h \psi^i}&=& \frac{3}{16}Z_{\psi}\bar{\psi}^i\slashed{p}\\
V_{\rm kin}^{\psi^{i\,T} h}&=& \frac{3}{16}Z_{\psi}\slashed{p}^T \bar{\psi}^{i\, T}\\
V_{\rm kin}^{\bar{\psi^i}h}&=& \frac{3}{16}Z_{\psi}\slashed{p}\psi^i,
\end{eqnarray}
where the momentum is always the momentum of the incoming graviton. 

The corresponding vertices with the TT mode vanish, as the
first term in \eqref{varkin} contains only the trace
mode, the second term vanishes by transversality for
constant external fermion fields and the last term vanishes
as the contraction $\gamma^{\mu}[\gamma^{\nu},
\gamma^{\kappa}]D_{\kappa}h^{\mathrm{TT}}_{\mu \nu}=0$.

The second variation of the kinetic fermion term with
respect to the metric contains only a TT contribution, as
the trace contribution is always of the form
$\bar{\psi}^i h \slashed{D}h \psi^i$, which can be rewritten
as a total derivative for constant external fermions. 

From the fact that we have constant external fermions at
 least one of the variations has to hit the covariant
derivative $\nabla_{\mu}$ and hence produce a $[\gamma^a,
\gamma^b]\delta \omega_{\mu\, a b}$. Accordingly the second
variation will necessarily contain $\gamma^{\mu}
[\gamma^{\kappa}, \gamma^{\lambda}]$. As there is one
derivative in the kinetic term, the vertex has to be
proportional to the momentum of one of the gravitons. The
only possibly structure that cannot be rewritten into a
total derivative is then $\gamma^{\mu}[\gamma^{\kappa},
\gamma^{\lambda}] h_{\kappa
\sigma}D_{\mu}h_{\lambda}^{\sigma}$. Our explicit
calculation now only has to fix the sign and the numerical
factor of the vertex. 
From \eqref{omega2var} and \eqref{omegavar} we deduce that
\begin{equation}
 \delta^2 \Gamma_{\rm kin}= i Z_{\psi} \int d^4 x 
\sqrt{g}\bar{\psi}^i\left(\frac{-1}{16}\right)\left(h^{\mu}_
{\lambda}\gamma^{\nu}[\gamma^{\lambda},
\gamma^{\kappa}]D_{\nu}h_{\kappa \mu}\right)\psi^i.
\end{equation}

The vertex that results from this expression is given by
\begin{eqnarray}
V^{h^{\mathrm{TT}}h^{\mathrm{TT}}}_{\text{kin}\, \mu \nu \kappa \lambda}&=& \frac{-1}{128}
p_{\tau}\bar{\psi}
 [\gamma^{\rho},\gamma^{\alpha}]\gamma^{\tau}\psi
\Bigl(\delta_{\mu\rho}\delta_{\kappa \alpha} \delta_{\nu
\lambda}+\delta_{\mu 
\rho}\delta_{\kappa\nu} \delta_{\alpha \lambda}\nonumber\\
&{}&+\delta_{\mu \lambda}\delta_{\nu \rho}\delta_{\kappa
\alpha}
+\delta_{\mu \kappa}\delta_{\nu \rho}\delta_{\alpha
\lambda}-\delta_{\rho \kappa}\delta_{\lambda
\nu}\delta_{\alpha \mu}\nonumber\\
&{}&-\delta_{\rho \kappa}\delta_{\lambda \mu}\delta_{\alpha\nu}-\delta_{\rho
  \lambda}\delta_{\kappa \nu}\delta_{\alpha \mu}-\delta_{\rho
  \lambda}\delta_{\kappa \mu}\delta_{\alpha \nu}\Bigr). 
\end{eqnarray}
The variations of the four-fermion term with respect to
 the metric are very simple: Due to
\begin{equation}
 \delta(\gamma^{\mu}\gamma_{\mu})= \delta(4)=0,
\end{equation}
only the determinant factor can contribute, and not the various $\gamma$
matrices. They always appear with completely contracted spacetime indices,
such that the above identity applies.  Hence the vertices containing three
external (anti)-fermions, one internal (anti)-fermion and one internal
graviton only exist for the trace mode, as $\delta \sqrt{g}=
\frac{1}{2}\sqrt{g}h$.  The vertices are given by
\begin{widetext}
\begin{eqnarray}
 V_{4f}^{h \bar{\psi}^{j\,T}}&=&- \frac{\bar{\lambda}_- + \bar{\lambda}_+}{2}\left(\bar{\psi}^i \gamma^{\mu}\psi^i \right)\psi^{j\, T}\gamma_{\mu}^T- \frac{\bar{\lambda}_- - \bar{\lambda}_+}{2}\left(\bar{\psi}^i \gamma^{\mu}\gamma^5\psi^i \right)\psi^{j\, T}\gamma^{5T}\gamma_{\mu}^T\nonumber\\
V_{4f}^{h \psi^j}&=&\frac{\bar{\lambda}_-+\bar{\lambda}_+}{2} \left(\bar{\psi}^i \gamma^{\mu}\psi^i \right) \bar{\psi}^j \gamma_{\mu} +\frac{\bar{\lambda}_- -\bar{\lambda}_+}{2} \left(\bar{\psi}^i \gamma^{\mu}\gamma^5\psi^i \right) \bar{\psi}^j \gamma_{\mu}\gamma^5\nonumber\\
V_{4f}^{\psi^{j\,T} h}&=& -\frac{\bar{\lambda}_-+\bar{\lambda}_+}{2}\left(\bar{\psi}^i \gamma^{\mu}\psi^i \right) \gamma_{\mu}^T\bar{\psi}^{j\, T}-\frac{\bar{\lambda}_--\bar{\lambda}_+}{2}\left(\bar{\psi}^i \gamma^{\mu}\gamma^5\psi^i \right) \gamma^{5T}\gamma_{\mu}^T\bar{\psi}^{j\, T}\nonumber\\
V_{4f}^{\bar{\psi^j}h}&=&\frac{\bar{\lambda}_-+\bar{\lambda}_+}{2}\left(\bar{\psi}^i \gamma^{\mu}\psi^i \right) \gamma_{\mu} \psi^j+\frac{\bar{\lambda}_--\bar{\lambda}_+}{2}\left(\bar{\psi}^i \gamma^{\mu}\gamma^5\psi^i \right) \gamma_{\mu} \gamma^5\psi^j.
\end{eqnarray}
%
%
The tadpole receives contributions from both the TT and 
the trace mode. The corresponding vertices are given by
\begin{equation}
 V_{4f}^{hh}=\frac{1}{16}\left(\bar{\lambda}_- (V-A)+\bar{\lambda}_+(V+A)
 \right), \quad
V_{4f\, \mu \nu \kappa \lambda}^{h^{\mathrm{TT}} h^{\mathrm{TT}}}=
-\frac{1}{8}\left(\delta_{\mu \kappa}\delta_{\nu \lambda}+\delta_{\mu
  \lambda}\delta_{\nu \kappa} \right)\left(\bar{\lambda}_-
(V-A)+\bar{\lambda}_+ (V+A) \right). 
\end{equation}
The variations of the four-fermion terms with respect to the fermions read as follows:
\begin{eqnarray}
 V_{4f}^{\psi^{i\,T} \bar{\psi}^{j\,T}}&=& \frac{\bar{\lambda}_-+\bar{\lambda}_+}{2}\left[2 \left(\gamma^{\mu\, T}\bar{\psi}^{i\, T} \right)\left(\psi^{j\, T}\gamma^{\mu\, T}\right)-2 \gamma^{\mu\, T}\delta^{ij}\left(\bar{\psi}^k \gamma^{\mu}\psi^k \right)\right]\nonumber\\
&{}&+\frac{\bar{\lambda}_--\bar{\lambda}_+}{2}\left[2 \left(\gamma^5\gamma^{\mu\, T}\bar{\psi}^{i\, T} \right)\left(\psi^{j\, T}\gamma^5\gamma^{\mu\, T}\right)-2\gamma^5 \gamma^{\mu\, T}\delta^{ij}\left(\bar{\psi}^k \gamma^5\gamma^{\mu}\psi^k \right)\right]\nonumber\\
V_{4f}^{\psi^{i\,T} \psi^j}&=& -\frac{\bar{\lambda}_-+\bar{\lambda}_+}{2}2 \left(\gamma^{\mu\, T}\bar{\psi}^{i\, T} \right)\left(\bar{\psi}^j \gamma^{\mu}\right)-\frac{\bar{\lambda}_--\bar{\lambda}_+}{2}2 \left(\gamma^5\gamma^{\mu\, T}\bar{\psi}^{i\, T} \right)\left(\bar{\psi}^j \gamma^{\mu}\gamma^5\right)\nonumber\\
V_{4f}^{\bar{\psi}^{i}\bar{\psi}^{j\,T}}&=& -\frac{\bar{\lambda}_-+\bar{\lambda}_+}{2}2 \left(\gamma^{\mu}\psi^i \right)\left(\psi^{j\, T}\gamma^{\mu \, T} \right)-\frac{\bar{\lambda}_--\bar{\lambda}_+}{2}2 \left(\gamma^{\mu}\gamma^5\psi^i \right)\left(\psi^{j\, T}\gamma^5\gamma^{\mu \, T} \right)\nonumber\\
V_{4f}^{\bar{\psi}^{i}\psi^j}&=& \frac{\bar{\lambda}_-+\bar{\lambda}_+}{2}\left[2 \delta^{ij}\gamma^{\mu}\left( \bar{\psi}^k \gamma^{\mu}\psi^k\right)+2 \left(\gamma^{\mu}\psi^i \right)\left(\bar{\psi}^j\gamma^{\mu} \right)\right]\nonumber\\
&{}&+\frac{\bar{\lambda}_--\bar{\lambda}_+}{2}\left[2 \delta^{ij}\gamma^{\mu}\gamma^5\left( \bar{\psi}^k \gamma^{\mu}\gamma^5\psi^k\right)+2 \left(\gamma^{\mu}\gamma^5\psi^i \right)\left(\bar{\psi}^j\gamma^{\mu}\gamma^5 \right)\right].
\end{eqnarray}
Here we suppress the Dirac index structure; by round brackets 
we indicate the way in which the Dirac indices of the terms 
have to be contracted.

\end{widetext}

\end{appendix}

\end{document}